\documentclass[aps,prc,showpacs,twocolumn,superscriptaddress,nofootinbib,preprintnumbers,floatfix]{revtex4-1}
\usepackage{hyperref}
\hypersetup{colorlinks=true,linkcolor=purple,anchorcolor=blue,citecolor=green, filecolor=blue,urlcolor=red,bookmarksnumbered=true,
	pdfview=FitB
}
\usepackage{color}
\usepackage{xcolor}
\usepackage{ulem}
\usepackage{mathrsfs,slashed,comment}
\usepackage{csquotes}
\colorlet{purple1}{blue!70!red}
\colorlet{darkred}{red!50!black}

\usepackage{graphicx}
\usepackage[bf,SL,BF]{subfigure}
\usepackage{psfrag}
\usepackage{color}
\usepackage{amssymb}
\usepackage{amsmath}
\usepackage{epstopdf}
\usepackage{natbib}

\usepackage{float}

\begin{document}
 \title{Analytical Evaluation of Elastic Lepton-Proton Two-Photon Exchange \\ in 
 Chiral Perturbation Theory}
\author{Poonam Choudhary}
     \email[]{poonamch@iitk.ac.in}
      \affiliation{Department of Physics, Indian Institute of Technology Kanpur, 
                  Kanpur-208016, India.}
\author{Udit Raha}
     \email[]{udit.raha@iitg.ac.in}
      \affiliation{Department of Physics, Indian Institute of Technology Guwahati, 
                  Guwahati - 781039, India.}
     \affiliation{Institute of Nuclear and Particle Physics, 
                  Ohio University, Athens, Ohio - 45701, USA }
\author{Fred Myhrer}
     \email[]{myhrer@mailbox.sc.edu}
      \affiliation{Department of Physics and Astronomy, 
                  University of South Carolina, Columbia, SC 29208, USA.}
 \author{Dipankar Chakrabarti}
     \email[]{dipankar@iitk.ac.in}
      \affiliation{Department of Physics, Indian Institute of Technology Kanpur, 
                  Kanpur-208016, India.}
\begin{abstract}
We present an exact evaluation of the two-photon exchange contribution to the elastic lepton-proton 
scattering process at low-energies using heavy baryon chiral perturbation theory. The evaluation is 
performed including next-to-leading order accuracy. This exact analytical evaluation contains all   
soft and hard two-photon exchanges and we identify the contributions missing in a soft-photon 
approximation approach. We evaluate the infrared divergent four-point box diagrams analytically 
using dimensional regularization. We also emphasize the differences between muon-proton 
and electron-proton scatterings relevant to the MUSE kinematics due to lepton mass differences.
\end{abstract}

 \maketitle
 
\section{INTRODUCTION}\label{sec:Introduction}
Scattering of light leptons off hadron targets has been the most time-honored precision tool to study 
the internal composite structure and dynamics of hadrons since the pioneering work of Hofstadter and 
McAllister~\cite{Hofstadter:1955ae}. The point-like nature of the leptons makes them ideal probes of the 
hadronic electromagnetic structure. Despite a century-long endeavor to understand the basic nucleon 
structure, fundamental gaps in our knowledge still persist. 

An accurate determination of the proton's electromagnetic form factors and parton distributions is known 
to shed much light on the constituent spin, charge, and magnetic distributions. However, various
systematic analyses of high-precision lepton-proton ($\ell^\pm$p) elastic scattering data, providing the 
cleanest possible information on the proton's internal structure, have brought forth several discrepancies
in the recent past which questions our conventional notion regarding the proton's structure as revealed
from the standard treatments of QED and QCD. A well-known discrepancy is the stark difference in the 
measured value of the proton's electric ($G^p_E$) to magnetic ($G^p_M$) form factor ratio ($G^p_E/G^p_M$)
at momentum transfers $Q^2$ beyond $\gtrsim 1$~(GeV/c)${}^2$ between two different popular experimental 
methodologies, namely, the {\it Rosenbluth Separation}~\cite{Rosenbluth:1950yq} and {\it Recoil 
Polarization Transfer}~\cite{Akhiezer:1974em,Arnold:1980zj,Gayou:2001qt} techniques (also see,
Refs.~\cite{Jones:1999rz,Perdrisat:2006hj,Punjabi:2015bba,Puckett:2010} for more details). A resolution 
to this ``form factor puzzle'' necessitates a closer investigation of the so-called {\it Two-Photon 
Exchange} (TPE) contributions to the radiative corrections to the elastic $\ell^\pm$p cross-section (for 
prominent past works and reviews, see e.g.,
Refs.~\cite{Arrington:2003,Guichon:2003,Blunden:2003sp,Rekalo:2004wa,Blunden:2005ew,Carlson:2007sp,Arrington:2011}). 
The TPE corrections give an additional higher-order contribution to the well-known leading order (LO) Born 
approximation contribution arising from the {\it One-Photon Exchange} (OPE) diagram which was assumed to 
dominate this electromagnetic scattering process at small momentum transfers. 

Likewise, there exists yet another puzzling scenario in regard to low-momentum transfers where the TPE 
consideration may prove to be a crucial game changer. This concerns the proton's charge radius, as obtained
from the slope of $G^p_E$ at $Q^2=0$. There exist two exclusive means to determine the charge radius, namely,
via scattering processes and via atomic spectroscopy. In particular, the muonic Lamb-shift measurements of
the {\it rms} charge radius by the CREMA Collaboration~\cite{Pohl:2010zza,Pohl:2013} are strikingly 
inconsistent with the prior CODATA recommended value ~\cite{Mohr:2012tt}. Such a discrepancy, the so-called 
``proton radius puzzle", has been an agenda of serious scientific contention over the last decade since its 
inception in 2013~\cite{Antognini:1900ns,Bernauer:2014,Carlson:2015,Bernauer:2020ont}. Despite the flurry of 
ingenious ideas and techniques introduced to fix the conundrum, the resolution of this discrepancy remained
unsettled thus far. We refer the reader to the recent status report as presented in Ref.~\cite{Gao:2021sml}. 
With no apparent fundamental flaws either conceptually or in the measurement process to explain this 
incongruity, the TPE processes may be singularly implicated as culpable under circumstantial evidence, 
vis-a-vis, the form factor puzzle~\cite{Arrington:2003,Guichon:2003,Blunden:2003sp,Rekalo:2004wa,Blunden:2005ew,Carlson:2007sp,Arrington:2011}.
It is conceivable that a rigorous evaluation of the TPE effects could potentially resolve both the form factor 
and radius discrepancies. Given the growing consensus in this regard, many new theoretical works on TPE 
studies have recently appeared in the
literature~\cite{Kivel:2012vs,Lorenz:2014yda,Tomalak:2014sva,Tomalak:2014dja,Tomalak:2015aoa,Tomalak:2015hva,Tomalak:2016vbf,Tomalak:2017npu,Koshchii:2017dzr,Tomalak:2018jak,Talukdar:2019dko,Peset:2021iul,Talukdar:2020aui,Kaiser:2022pso,Guo:2022kfo}.

Of the several newly commissioned high-precision scattering experiments~\cite{Bernauer:2020ont}, the ongoing MUSE
Collaboration project at PSI~\cite{Gilman:2013eiv} is one such endeavor, uniquely designed to simultaneously scatter 
leptons ($\ell^-\equiv e^-,\,\mu^-$) as well as anti-leptons ($\ell^+\equiv e^+,\,\mu^+$) off a proton target. One  
specialty of MUSE will be its uniqueness in pinning down the {\it charge-odd} contributions to the unpolarized 
lepton-proton ($\ell^\pm$p $\to \ell^\pm$p) scattering, which arguably includes the TPE as the dominant process.
Nevertheless, isolating the charge-odd contributions does not necessarily preclude other competing low-energy 
chiral-radiative contributions~\cite{Talukdar:2020aui} in affecting the extraction of the pure TPE loop contributions. 
In particular, the recent estimation~\cite{Talukdar:2020aui} of the ``soft'' bremsstrahlung  
($\ell^\pm$p $\to \ell^\pm$p$\,\gamma^*_{\rm soft}$) corrections to {\it next-to-leading} order (NLO) in {\it heavy 
baryon chiral perturbation theory} 
(HB$\chi$PT)~\cite{Gasser:1982ap,Jenkins:1990jv,Bernard:1992qa,Ecker:1994pi,Bernard:1995dp} revealed novel chiral-odd 
constituents large enough (and of opposite sign) to supersede the TPE effects pertinent to the MUSE 
kinematics.\footnote{Notably this observation contrasts the standard expectation based on relativistic hadronic models,
namely, that the TPE loop diagrams constitute the {\it only} charge-odd contribution responsible for asymmetries between
the lepton and anti-lepton scatterings. All other virtual radiative effects (vacuum polarization, vertex, and self-energy 
corrections) are known to be charge symmetric.} Given that the TPE processes are major sources of systematic uncertainty, 
their accurate theoretical estimation is of pivotal interest for the purpose of precision analysis of the future MUSE 
data aimed at sub-percentage accuracy. 

Since the inception of the radius puzzle, the importance of low-energy TPE contributions has been extensively explored 
{\it via} diverse approaches such as QED-inspired hadronic 
models~\cite{Tomalak:2014dja,Tomalak:2015aoa,Tomalak:2015hva,Koshchii:2017dzr,Guo:2022kfo}, Dispersion
techniques~\cite{Lorenz:2014yda,Tomalak:2014sva,Tomalak:2016vbf,Tomalak:2017npu,Tomalak:2018jak,Guo:2022kfo}, and 
Effective Field Theories (EFTs) such as Non-Relativistic Quantum Electrodynamics (NRQED)~\cite{Peset:2021iul}, 
Baryon Chiral Perturbation Theory (B$\chi$PT)~\cite{Cao:2021nhm} and HB$\chi$PT~\cite{Talukdar:2019dko,Talukdar:2020aui}. 
It is well-known that restricting to the low-energy regime, the TPE with elastic proton intermediate state yields the 
dominant contribution to the radiative corrections~\cite{Blunden:2003sp}. Besides the proton, the inclusion of other 
excited nucleon intermediate states, such as the spin-isospin quartet of $\Delta (1232)$ resonances, can yield interesting
results~\cite{Bernauer:2014} relevant to the MUSE kinematics that leads to a better understanding of the non-perturbative 
aspects of the TPE contributions~\cite{Kondratyuk:2005kk}. The TPE loop diagrams which have been evaluated in the past for
a wide range of intermediate to high momentum transfers~\cite{Christy:2021snt} are either known to be model-dependent or 
used unwarranted simplifications (generally referred to as {\it soft photon approximation} or SPA \cite{Tsai:1961zz}) in 
the treatment of the intricate four-point functions, (see e.g., Refs.~\cite{Koshchii:2017dzr,Tomalak:2014dja}). In this work,
we present the evaluation of the TPE loop contributions using HB$\chi$PT without taking recourse to SPA methods. 

The outline of the paper is as follows. In Sec.~\ref{TPE-diagrams}, after a brief discussion of the HB$\chi$PT Lagrangian 
needed for our intended accuracy, namely, including interactions that at {\it next-to-leading order} (NLO) in the power 
counting, we introduce the possible topologies of the TPE loop diagrams that contribute at LO and NLO. In 
Sec.~\ref{TPE-contri}, we discuss the amplitudes of these loop diagrams and their contributions to the elastic cross-section. 
In Sec.~\ref{Results}, we discuss our numerical results and compare them with other works. Finally, we present our summary and
conclusions in Sec.~\ref{Summary}. Appendix~\ref{appA} contains the notations of the various generic two-, three- and 
four-point integral functions that contribute to the TPE corrections to the cross-section. In this work, all such relevant 
integrals have been evaluated exactly, and their analytical expressions are collected in Appendix~\ref{appB}.

\section{TPE diagrams in HB$\chi$PT }\label{TPE-diagrams}
In this work, we evaluate the LO and NLO TPE contributions to lepton-proton elastic 
scattering cross-section using HB$\chi$PT, where only parts of the LO and NLO chiral Lagrangian contribute. 
In particular, the pion degrees of freedom  that appear only {\it via} the {\it next-to-next-to-leading} order 
(NNLO) loop diagrams are absent in our current accuracy. This means that in our Lagrangian the non-linear 
Goldstone field $u = \sqrt{U}$ effectively contributes as the identity matrix in two-dimensional isospin space, 
namely, $u\mapsto {\mathbb I}_{2\times 2}$. Consequently, the relevant LO and NLO parts of the chiral Lagrangian 
become\footnote{In ${\cal L}_{\pi N}^{(1)}$ the anomalous magnetic moment terms contribute at NNLO and are 
therefore not explicitly given in Eq.(2).}
\begin{equation}
	\mathcal{L}^{(0)}_{\pi N}=\overline{N} (iv\cdot D+g_A S\cdot u)N\,,
\label{eq:Lchiral0}
\end{equation}
and
\begin{equation}
	\mathcal{L}^{(1)}_{\pi N}=\overline{N} \left\{ \frac{1}{2M}(v\cdot D)^2-\frac{1}{2M}D\cdot D+...\right\} N\,,
	\label{eq:Lchiral1}
\end{equation}
respectively, where $N=(\rm{p\,\,n})^{\rm T}$ is the heavy nucleon spin-isospin doublet spinor field, $g_A=1.267$ 
is axial coupling constant of the nucleon, and $v_\mu$ and $S_\mu$ are the nucleon velocity and spin four-vectors 
satisfying the condition, $v\cdot S= 0$. Here, the standard choice is $v=(1,{\bf 0})$ such that 
$S=(0,\boldsymbol{\sigma}/2)$. Furthermore, the gauge covariant derivative is 
\begin{equation}
D_\mu = \partial_\mu+\Gamma_\mu-iv^{(s)}_\mu\,, 
\end{equation}
with the chiral connection given by
\begin{equation}
\Gamma_\mu = \frac{1}{2}[u^\dagger(\partial_\mu-ir_\mu)u+u(\partial_\mu-il_\mu)u^\dagger]\,,
\end{equation}
and the chiral vielbein is $u_\mu = iu^\dagger\nabla_\mu U u^\dagger$, where
\begin{equation}
	\quad \nabla_\mu U=\partial_\mu U-ir_\mu U+iUl_\mu\,.
\end{equation} 
Here, $l_\mu$ and $r_\mu$ are external iso-vector chiral source fields, which in our case are given by the photon 
field, namely, $r_\mu=l_\mu=-e\frac{\tau^3}{2}A_\mu$, with $\tau^3$ being the third Pauli isospin matrix. Finally, 
$v_\mu^{(s)}=-e\frac{I}{2}A_\mu$ is the external iso-scalar vector source field.

\begin{figure}
\begin{center}
			\includegraphics[width=0.44\textwidth]{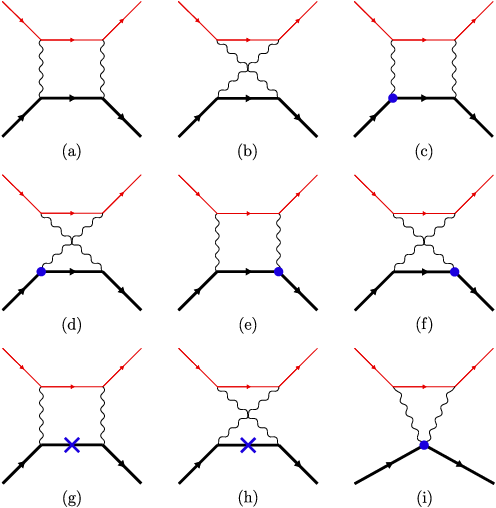}
			\caption{TPE diagrams of $\mathcal{O}(\alpha^2)$ which contribute to the $\mathcal{O}(\alpha^3)$ 
                     elastic cross-section. The thick, thin, and wiggly lines denote proton, lepton, and photon 
                     propagators, respectively. The blobs and crosses denote NLO insertions of the proton-photon
                     vertices and proton propagators, respectively. }
			\label{fig1}
\end{center}
\end{figure}

Figure~\ref{fig1} shows all relevant TPE diagrams up-to-and-including NLO chiral order.
The first two diagrams, namely, the box diagram (a) and the crossed-box diagram (b) are of LO where both the 
photon-proton vertices as well as the proton propagator stem from the LO Lagrangian. However, these two LO 
diagrams also contain ${\mathcal O}(1/M)$ parts originating from the proton propagators (see below) which are 
kinematically suppressed. In $\chi$PT the perturbative chiral expansion is determined in terms of powers of 
the expansion parameter $Q/\Lambda_\chi$, where $Q$ is a generic momentum scale of the process and 
$\Lambda_\chi\approx 1$~GeV/c is the breakdown scale of the theory. Since the value of $\Lambda_\chi$ is 
approximately equal to the proton's mass $M$, the HB$\chi$PT additionally incorporates a non-perturbative 
power or recoil correction of the interaction vertices and propagators {\it via} an expansion in powers of the
inverse proton's mass, namely, $Q/M\sim Q/ \Lambda_\chi$. It is also important to formally distinguish such 
{\it dynamical recoil} corrections from the naive {\it kinematical recoil} corrections. Thus, for instance, the 
difference in incoming lepton and outgoing lepton energies, $E - E^\prime =-Q^2/(2M)$, and the corresponding 
lepton velocities, $\beta-\beta^\prime=-Q^2(1-\beta^2)/(2ME\beta)+{\mathcal O}(1/M^2)$, are interpreted as 
kinematic recoil corrections of order $1/M$. Nevertheless, in practice, such corrections are also referred to 
as NLO corrections. Consequently, diagrams (a) and (b) are not ``true LO" chiral contributions since they contain
order $1/M$ ``NLO contributions'' from the proton propagator, e.g., 
$v\cdot p_p \approx -p^2_p/(2 M)+{\mathcal O}(1/M^2)$, where $p_p$ is a generic small off-shell four-momentum 
of the proton which is related to the corresponding full four-momentum in the heavy baryon formalism by the 
relation, $P^\mu=Mv^\mu+p^\mu_p$. Next, the diagrams (c), (d), (e), and (f) are genuine NLO chiral order diagrams
since they contain one NLO vertex, i.e., one photon-proton vertex is taken from NLO chiral Lagrangian whereas the
propagators in each of these diagrams are taken from LO chiral Lagrangian. Likewise, the diagrams (g) and (h) are 
also genuine NLO box and crossed-box diagrams where the proton propagator in each diagram is derived from NLO 
chiral Lagrangian whereas the proton-photon vertices are taken from LO chiral Lagrangian. Finally, the 
triangular-shaped seagull diagram (i) contains an NLO vertex of $\mathcal{O}(e^2)$ with no proton intermediate 
state. The explicit amplitude for each of the TPE diagrams is given by the following expressions:
\begin{widetext}
\begin{eqnarray}
		{\mathcal M}^{(a)}_{\rm box} \!&=&\! e^4\int \!\frac{{\rm d}^4k}{(2\pi)^4i}
		\frac{\left[\bar u(p^\prime)\gamma^\mu(\not{\!p}\,-\not{\!k}+\,m_l)\gamma^\nu u(p)\right]\,\left[\chi^\dagger(p_p^\prime)v_\mu 
            v_\nu \chi(p_p)\right]}{(k^2+i0)\,[(Q-k)^2+i0]\,(k^2-2k\cdot p+i0)\,(v\cdot k+v\cdot p_p+i0)}\,,
		\label{M_a}\\
\nonumber\\
		{\mathcal M}^{(b)}_{\rm xbox} \!&=&\! e^4\int \!\frac{{\rm d}^4k}{(2\pi)^4i}
		\frac{\left[\bar u(p^\prime)\gamma^\mu(\not{\!p}\,-\not{\!k}+\,m_l)\gamma^\nu u(p)\right]\,\left[\chi^\dagger(p_p^\prime)v_\mu 
            v_\nu \chi(p_p)\right]}{(k^2+i0)\,[(Q-k)^2+i0]\,(k^2-2k\cdot p+i0)\,(-v\cdot k+v\cdot p^{\prime}_p+i0)}\,,
		\label{M_b}\\
\nonumber\\
		{\mathcal M}^{(c)}_{\rm box} \!&=&\! \frac{e^4}{2M}\int \!\frac{{\rm d}^4k}{(2\pi)^4i}
		\frac{\left[\bar u(p^\prime)\gamma^\mu(\not{\!p}\,-\not{\!k}+\,m_l)\gamma^\nu u(p)\right]\,
	      \left[\chi^\dagger(p_p^\prime)\{v_\mu(k + 2p_p)_\nu-v_\mu v_\nu v\cdot (k + 2 p_p)\} 
            \chi(p_p)\right]}{(k^2+i0)\,[(Q-k)^2+i0]\,(k^2-2k\cdot p+i0)\, (v\cdot k + v\cdot p_p+i0)}\,,
		\label{M_c}\\
\nonumber\\
		{\mathcal M}^{(d)}_{\rm xbox} \!&=&\! \frac{e^4}{2M}\int \!\frac{{\rm d}^4k}{(2\pi)^4i}
		\frac{\left[\bar u(p^\prime)\gamma^\mu(\not{\!p}\,-\not{\!k}+\, m_l)\gamma^\nu u(p)\right]\, 
	      \left[\chi^\dagger(p_p^\prime)\{v_\nu(p_p + p_p^{\prime} - k)_\mu - v_\mu v_\nu v \cdot (p_p + p_p^{\prime} - k)\} 
           \chi(p_p)\right]}{(k^2+i0)\,[(Q-k)^2+i0]\,(k^2-2k\cdot p+i0)\,( v\cdot p^{\prime}_p-v\cdot k+i0)}\,,\qquad\,
		\label{M_d}\\
\nonumber\\
		{\mathcal M}^{(e)}_{\rm box} \!&=&\! \frac{e^4}{2M}\int \!\frac{{\rm d}^4k}{(2\pi)^4i}
		\frac{\left[\bar u(p^\prime)\gamma^\mu(\not{\!p}\,-\not{\!k}+\,m_l)\gamma^\nu u(p)\right]\,
		\left[\chi^\dagger(p_p^\prime)\{v_\nu(p_p + p_p^\prime + k)_\mu - v_\mu v_\nu v\cdot(p_p + p_p^\prime + k)\} 
            \chi(p_p)\right]}{(k^2+i0)\,[(Q-k)^2+i0]\,(k^2-2k\cdot p +i0)\,( v\cdot p_p+v\cdot k+i0)}\,,\qquad\,
		\label{M_e}\\
\nonumber\\
		{\mathcal M}^{(f)}_{\rm xbox} \!&=&\! \frac{e^4}{2M}\int \!\frac{{\rm d}^4k}{(2\pi)^4i}
		\frac{\left[\bar u(p^\prime)\gamma^\mu(\not{\!p}\,-\not{\!k}+\,m_l)\gamma^\nu u(p)\right]\,
		\left[\chi^\dagger(p_p^\prime)\{v_\mu(2p_p^\prime - k)_\nu - v_\mu v_\nu v\cdot(2p_p^\prime - k) \} 
        \chi(p_p)\right]}{(k^2+i0)\,[(Q-k)^2+i0]\,(k^2-2k\cdot p+i0)\,( v\cdot p^{\prime}_p -v\cdot k +i0)}\,,
		\label{M_f}\\
\nonumber\\ 
		\mathcal{M}^{(g)}_{\rm box} \!&=&\! \frac{e^4}{2M}\int \!\frac{{\rm d}^4k}{(2\pi)^4i}
		\frac{\left[\bar u(p^\prime)\gamma^\mu(\not{\!p}\,-\not{\!k}+\,m_l)\gamma^\nu u(p)\right]\,
		\left[\chi^\dagger(p_p^\prime)v_\mu v_\nu\chi(p_p)\right]}{(k^2+i0)\,[(Q-k)^2+i0]\,(k^2-2k\cdot p+i0)}
		\left(1-\frac{(p_p+k)^2}{( v\cdot p_p +v\cdot k+i 0)^2}\right)\,,
		\label{M_g}
\end{eqnarray}
\begin{eqnarray}
		{\mathcal M}^{(h)}_{\rm xbox} \!&=&\! \frac{e^4}{2M}\int \! \frac{{\rm d}^4k}{(2\pi)^4i}
		\frac{\left[\bar u(p^\prime)\gamma^\mu(\not{\!p}\,-\not{\!k}+\,m_l)\gamma^\nu u(p)\right]\,
		\left[\chi^\dagger(p_p^\prime)v_\mu v_\nu\chi(p_p)\right]}{(k^2+i0)\,[(Q-k)^2+i0]\,(k^2-2k\cdot p+i0)}
		\left(1-\frac{(p^{\prime}_p-k)^2}{ ( v\cdot p^{\prime}_p - v\cdot k+i 0)^2}\right)\,,
		\label{M_h}\\
\nonumber\\ 
		{\mathcal M}^{(i)}_{\rm seagull} \!&=&\! \frac{2 e^4}{2M}\int \!\frac{{\rm d}^4k}{(2\pi)^4i}
		\frac{\left[\bar u(p^\prime)\gamma^\mu(\not{\!p}\,-\not{\!k}+\,m_l)\gamma^\nu u(p)\right]\,
		\left[\chi^\dagger(p_p^\prime)(v_\mu v_\nu-g_{\mu \nu})\chi(p_p)\right]}{(k^2+i0)\,[(Q-k)^2+i0]\,(k^2-2k\cdot p+i0)}\,,
		\label{M_i}
\end{eqnarray}
\end{widetext}
where $u(p)$ and $\bar u(p^\prime)$ are the incoming and outgoing lepton Dirac spinors, and $\chi(p_p)$ and 
$\chi^\dagger(p_p^\prime)$ are the corresponding non-relativistic proton Pauli spinors. We have used the fact
that the proton propagators up-to-and-including NLO for the box and crossed-box TPE loops diagrams are 
respectively given as
\begin{widetext}
\begin{eqnarray}
     iS^{(1/M)}_{\rm full}(p_p+k) \!&=&\! 
     \frac{i}{v\cdot(p_p + k)+i0}+ \frac{i}{2M}\left[1-\frac{(p_p+k)^2}{(v\cdot p_p + v\cdot k+i0)^2}\right] + {\cal O}(M^{-2}) \,,  
     \nonumber\\
    iS^{(1/M)}_{\rm full}(p^\prime_p - k) \!&=&\! 
    \frac{i}{v\cdot(p^\prime_p - k)+i0}+ \frac{i}{2M}\left[1-\frac{(p^\prime_p - k)^2}{(v\cdot p^\prime_p 
    - v\cdot k+i0)^2}\right] + {\cal O}(M^{-2}) \,.
\end{eqnarray}
\end{widetext} 
 
Next, examining the possible cancellations of contributions at the amplitude level, the following comments are in
order: 
\begin{itemize}
\item First, it is convenient for us to use the {\it lab}-frame kinematics where the initial proton is at rest, i.e., 
$p_p=0$, which means that the four-momentum transfer is $Q=p^{\prime}_p-p_p=p^{\prime}_p $. 
\item Second, since we observe that $v\cdot Q = E - E^\prime=-Q^2/(2M)\sim {\mathcal O}(1/M)$, this difference can
therefore be neglected in the numerators of the NLO amplitudes 
$$\mathcal{M}^{(c)}_{\rm box}\,,\,\, \mathcal{M}^{(d)}_{\rm box}\,,\,\, \mathcal{M}^{(e)}_{\rm box}\,,\,\, \text{and} \,\,\, 
\mathcal{M}^{(f)}_{\rm xbox} \,,$$  
as they give rise to higher-order contributions. What remains in the numerators of these four amplitudes following 
the factor $v_\mu v_\nu$ are the terms $v\cdot k$, only. However, it is important to retain the $v\cdot Q$ terms in 
the denominators of the proton propagators for consistent evaluation of the loop integrals. In many cases, they serve 
as natural regulators against infrared (IR) divergences. Nevertheless, as stated, we may ignore all 
$v\cdot p_p^\prime = v\cdot Q$ terms in the NLO expressions after evaluations of the integrals, only. 
\end{itemize}
Consequently, when we sum these amplitudes, the terms containing the factors $ v_\mu v_\nu $ in the four NLO amplitudes 
of $\mathcal{M}^{(c)}_{\rm box}$, $\mathcal{M}^{(d)}_{\rm xbox}$, $\mathcal{M}^{(e)}_{\rm box}$ and 
$\mathcal{M}^{(f)}_{\rm xbox}$, cancel with the two $v_\mu v_\nu $ terms with coefficient 1 (within the braces) of  
$\mathcal{M}^{(g)}_{\rm box}$ and $\mathcal{M}^{(h)}_{\rm xbox}$, plus the $v_\mu v_\nu$ term in the seagull 
amplitude $\mathcal{M}^{(i)}_{\rm seagull}$. After such partial cancellations between the NLO amplitudes, the remaining
parts of the NLO box diagrams (c), (e), and (g), and the seagull diagram (i), along with the unaltered LO amplitude (a) 
are given in the {\it lab}-frame as follows:
\begin{widetext}
\begin{eqnarray} 
	  i{\mathcal M}^{(a)}_{\rm box} \!&=&\! e^4\int \frac{{\rm d}^4k}{(2\pi)^4}
	    \frac{\left[\bar u(p^\prime)\gamma^\mu(\not{\!p}-\not{\!k}+\,m_l)\gamma^\nu u(p)\right]\,
        \left[\chi^\dagger(p_p^\prime)v_\mu v_\nu \chi(p_p)\right]}{(k^2+i0)\,[(Q-k)^2+i0]\,(k^2-2k\cdot p+i0)\,(v\cdot k+i0)}\,,
        \nonumber \\
	i\widetilde{\mathcal M}^{(c)}_{\rm box} \!&=&\! \frac{e^4}{2M}\int \frac{{\rm d}^4 k}{(2 \pi)^4} 
        \frac{\left[\bar u(p^\prime)\gamma^\mu(\not{\!p}-\not{\!k}+\,m_l)\gamma^\nu u(p)\right]\,
        \left[\chi^{\dagger}(p^\prime_p){v_\mu k_\nu} \chi(p_p)\right]}{(k^2+i0)\,[(Q-k)^2+i0]\,(k^2-2 k\cdot p+i0)\,(v\cdot k+i0)}\,,
        \nonumber \\
	i\widetilde{\mathcal M}^{(e)}_{\rm box} \!&=&\! \frac{e^4}{2M}\int \frac{{\rm d}^4 k}{(2 \pi)^4} 
        \frac{\left[\bar u(p^\prime)\gamma^\mu(\not{\!p}-\not{\!k}+\,m_l)\gamma^\nu u(p)\right]\,
        \left[\chi^{\dagger}(p^\prime_p){v_\nu (k+Q)_\mu} \chi(p_p)\right]}{(k^2+i0)\,[(Q-k)^2+i0]\,(k^2-2 k\cdot p+i0)\,(v\cdot k+i0)}\,,
        \nonumber \\
	i\widetilde{\mathcal M}^{(g)}_{\rm box} \!&=&\! \frac{e^4}{2M}\int \frac{{\rm d}^4 k}{{2 \pi}^4} 
        \frac{\left[\bar u(p^\prime)\gamma^\mu(\not{\!p}-\not{\!k}+\,m_l)\gamma^\nu u(p)\right]\,
        \left[\chi^{\dagger}(p^\prime_p) v_\mu v_\nu \chi(p_p)\right]}{(k^2+i0)\,[(Q-k)^2+i0]\,(k^2-2 k\cdot p+i0)} 
        \Bigg(-\frac{ k^2}{(v\cdot k+i 0)^2}\Bigg)
        \nonumber \\      
	    &=&\! -\,\frac{e^4}{2M}\int \frac{{\rm d}^4 k}{{2\pi}^4} 
        \frac{\left[\bar u(p^\prime)\gamma^\mu(\not{\!p}-\not{\!k}+\,m_l)\gamma^\nu u(p)\right]\,
        \left[\chi^{\dagger}(p^\prime_p) v_\mu v_\nu \chi(p_p)\right]}{[(Q-k)^2+i0]\,(k^2-2 k\cdot p+i0)\,(v\cdot k+i0)^2}\,,
        \nonumber \\
    i\widetilde{\mathcal M}^{(i)}_{\rm seagull} \!&=&\! -\,\frac{e^4}{M}\!\!\int \!\!\frac{{\rm d}^4k}{(2 \pi)^4}
	    \frac{\left[\bar u(p^\prime)\gamma^\mu(\not{\!p}-\not{\!k}+\,m_l)\gamma_\mu u(p)\right]\,
	    \left[\chi^\dagger(p_p^\prime)\,\chi(p_p)\right]}{(k^2+i0)\,[(Q-k)^2+i0]\,(k^2-2k\cdot p+i0)}\,. 
\label{17} 
\end{eqnarray}
Regarding the other amplitudes for the crossed-box diagrams, it is convenient to shift the integration variable $k$ by
means of the transformation $k\rightarrow -k+Q$, which yields the following expressions:
\begin{eqnarray}
	i{\mathcal M}^{(b)}_{\rm xbox} \!&=&\! e^4\int \frac{{\rm d}^4k}{(2\pi)^4}
	    \frac{\left[\bar u(p^\prime)\gamma^\mu(\not{\!p}+\not{\!k}-\not{\!Q}+\,m_l)\gamma^\nu u(p)\right]\,
        \left[\chi^\dagger(p_p^\prime)v_\mu v_\nu \chi(p_p)\right]}{(k^2+i0)\,[(Q-k)^2+i0]\,(k^2+2k\cdot p^\prime+i0)\,(v\cdot k+i0)}\,,
        \nonumber\\
	i\widetilde{\mathcal M}^{(d)}_{\rm xbox}\!&=&\! \frac{e^4}{2M}\int \frac{{\rm d}^4 k}{(2 \pi)^4} 
        \frac{\left[\bar u(p^\prime)\gamma^\mu(\not{\!p}+\not{\!k}-\not{\!Q}+\,m_l)\gamma^\nu u(p)\right]\,
        \left[\chi^{\dagger}(p^\prime_p){v_\nu k_\mu} \chi(p_p)\right]}{(k^2+i0)\,[(Q-k)^2+i0]\,(k^2+2 k\cdot p^\prime +i0)\,(v\cdot k+i0)}\,,
        \nonumber\\
	i\widetilde{\mathcal M}^{(f)}_{\rm xbox}\!&=&\! \frac{e^4}{2M}\int \frac{{\rm d}^4 k}{(2 \pi)^4} 
        \frac{\left[\bar u(p^\prime)\gamma^\mu(\not{\!p}+\not{\!k}-\not{\!Q}+\,m_l)\gamma^\nu u(p)\right]\,
        \left[\chi^{\dagger}(p^\prime_p){v_\mu (k+Q)_\nu} \chi(p_p)\right]}{(k^2+i0)\,[(Q-k)^2+i0]\,(k^2+2 k\cdot p^\prime +i0)\,(v\cdot k+i0)}\,,
        \nonumber\\
	i\widetilde{\mathcal M}^{(h)}_{\rm xbox}\!&=&\! -\,\frac{e^4}{2M}\int \frac{{\rm d}^4 k}{{2 \pi}^4} 
        \frac{\left[\bar u(p^\prime)\gamma^\mu(\not{\!p}+\not{\!k}-\not{\!Q}+\,m_l)\gamma^\nu u(p)\right]\,
        \left[\chi^{\dagger}(p^\prime_p) v_\mu v_\nu \chi(p_p)\right]}{[(k-Q)^2+i0]\,(k^2+2 k\cdot p^\prime +i0)\,(v\cdot k+i0)^2}\,.	
\end{eqnarray}
\end{widetext}
The above TPE amplitudes include only LO and NLO amplitude contributions, i.e., we neglect all higher-order 
contributions. These diagrams constitute the radiative corrections to the LO Born amplitude and contribute
to the elastic unpolarized lepton-proton scattering cross-section. Their contribution to the cross-section 
is obtained {\it via} the interference of these TPE amplitudes with the LO Born 
diagram~\cite{Talukdar:2019dko,Talukdar:2020aui}. In the following section, we present the complete analytical
evaluation of the above TPE contributions to the cross-section up-to-and-including NLO in HB$\chi$PT without 
resorting to any kind of approximation other than the usual perturbative NLO truncation. For this
purpose, we first need to isolate the finite parts of the TPE corrections from the IR-divergent parts which 
are unphysical. 

\section{TPE contributions to the elastic cross-section}
\label{TPE-contri} 
The differential cross-section due to the TPE is given by the fractional TPE contribution 
$\delta^{\rm (box)}_{\gamma\gamma} (Q^2)\sim \mathcal{O} (\alpha)$ times the LO Born differential cross-section 
[i.e., of ${\mathcal O}(\alpha^2)$], namely,
\begin{eqnarray}
\Big[\frac{{\rm d} \sigma_{el}(Q^2)}{{\rm d} \Omega^\prime_l}\Big]_{\gamma \gamma}
=\Big[\frac{{\rm d} \sigma_{el}(Q^2)}{{\rm d} \Omega^\prime_l}\Big]_{\gamma}\delta^{\rm (box)}_{\gamma\gamma} (Q^2)\,.
\label{eq:crosssection}
\end{eqnarray} 
For unpolarized elastic lepton-proton scattering cross-section only the finite real part of the amplitude 
contributes:
\begin{eqnarray} 
\delta^{\rm (box)}_{\gamma\gamma} (Q^2)=\frac{2{\mathcal R}e\sum\limits_{spins}
\left[{\mathcal M}^{(0)*}_\gamma\,{\mathcal M}^{\rm (box)}_{\gamma\gamma}\right]}{\sum\limits_{spins}
\left|{\mathcal M}^{(0)}_\gamma\right|^2}-\delta^{\rm (box)}_{\rm IR}(Q^2)\,,\quad\,\,
\label{DELTABOX}
\end{eqnarray}
where ${\mathcal M}_\gamma $ corresponds to the LO OPE or the Born amplitude for the elastic lepton-proton 
scattering process, and is given by\,\footnote{We note that there are additional non-vanishing contributions to 
the TPE cross-section arising from the interference of the proton's spin-independent NLO Born 
amplitude~\cite{Talukdar:2020aui}:
\begin{eqnarray*}
\mathcal{M}^{(1)}_\gamma \!&=&\! -\,\frac{e^2}{2 M Q^2}[{\bar u}(p^\prime)\gamma^\mu\,u(p)]\,
\Big[\chi^\dagger(p_p^\prime)
\nonumber\\
&&\qquad \times\, \left\{(p_p + p_p^\prime)_\mu - v_\mu v \cdot (p_p + p_p^\prime) \right\} \chi(p_p)\Big]\,,
\end{eqnarray*}
with the LO TPE amplitudes, ${\mathcal M}^{(a)}_{\rm box}$ and  ${\mathcal M}^{(b)}_{\rm xbox}$. However, these
contributions are of ${\mathcal O}(1/M^2)$ and hence ignored in this work. The corresponding contribution from 
the spin-dependent NLO Born amplitude (see Ref.~\cite{Talukdar:2020aui}) identically vanishes in this case.}
\begin{eqnarray}
    {\mathcal M}^{(0)}_\gamma=-\frac{e^2}{Q^2} \big[\bar{u}(p^\prime)\gamma^\mu u(p) \big] 
    \left[\chi^{\dagger}(p^\prime_p) v_\mu \chi(p_p)\right]\,.
\end{eqnarray} 
The term $\delta^{\rm (box)}_{\rm IR}(Q^2)$ [see Eq.~\eqref{deltaIRbox}] collects all the IR-divergent parts of the
TPE amplitudes and is expected to cancel exactly with IR-divergent terms of the soft-bremsstrahlung counterparts. 
The TPE amplitude, ${\mathcal M}^{\rm (box)}_{\gamma \gamma}$ is the sum of the box,
crossed-box, and seagull diagrams presented in Fig.~\ref{fig1}. As we calculate the LO and NLO contributions to the 
elastic cross-section, i.e., of $\mathcal O(\alpha^3/M)$, from each of the aforementioned TPE box and crossed-box 
diagrams, we  retain all possible $\mathcal O(1/M)$ terms. As mentioned earlier, since the LO TPE box (a) and the 
crossed box (b) diagrams have both LO and NLO terms, in order to derive the ``true LO" contribution in the theory, it 
is necessary to eliminate the ${\mathcal O}(1/M)$ contributions from these LO diagrams.

In our HB$\chi$PT calculation of the TPE box (and crossed-box) amplitudes, there arise four-point integrals where the 
proton propagator is non-relativistic and linear in loop momentum. These integrals are simplified using a modified 
form of Feynman parametrization~\cite{zupan:2002}. The calculation of such integrals differs from the standard approach 
of evaluating loop diagrams with relativistic proton propagators. In this work, we derive the expressions for the TPE 
four-point functions in terms of two- three-, and four-point scalar {\it master integrals} using successive 
{\it integration-by-parts} (IBP) techniques, combining with decomposition {\it via} the method of partial 
fractions~\cite{Chetyrkin:1981qh}. The IBP method is a widely known technique whereby complicated loop-integrals 
containing products of four or more propagators are decomposed in terms of known simpler master integrals up to three 
propagators. The master integrals, in turn, are straightforward to evaluate using standard techniques for evaluating 
Feynman loop integrals. However, such methodologies are primarily meant to tackle ultraviolet (UV) divergences in loop 
functions with relativistic propagators that are quadratic in loop four-momentum. In the current case where 
IR divergences are involved in loop functions with the non-relativistic propagators, the application of such existing 
techniques becomes less obvious. In Ref.~\cite{zupan:2002}, Zupan demonstrated the evaluation of one-loop scalar 
integrals up to four-point functions with heavy quark propagators (also linear in four-momentum). The propagator 
structure of heavy quark is very similar to the heavy baryon (proton) propagator in HB$\chi$PT. In effect, the HB$\chi$PT
theory was formulated on similar lines following the ideas of Heavy Quark Effective Theory (HQET), (see, e.g., 
Ref.~\cite{Grozin:2000cm}). Therefore, it is in principle straightforward to extend Zupan's method~\cite{zupan:2002} to 
evaluate the TPE integrals in our HB$\chi$PT calculations. One should, however, maintain some degree of caution in 
directly using the results of Ref.~\cite{zupan:2002}, since the techniques presented therein were meant to handle either 
finite or UV-divergent two-, three- and four-point loop functions with massive propagators. The integrals appearing in 
our case are in contrast IR-divergent-containing photon propagators. In that case, it may seem straightforward to 
introduce photon mass regulators to evaluate such IR-divergent integrals. However, such a mass cut-off regularization 
scheme to evaluate the IR-divergent Feynman integrals using Zupan's technique does not {\it a priori} distinguish between 
$v \cdot k+ i0$ and $ v \cdot k -i0$ in the propagator denominators with four-momentum $k^\mu$, and may lead to 
discrepancies in estimating the correct analytic structure of the Feynman amplitudes. In contrast, by employing dimensional
regularization to tackle the IR-divergent integrals using methods of complex analysis, albeit much more involved, one can 
avoid such ambiguities in evaluating the loop amplitudes.

\subsection{Chiral order contributions and the ${\mathcal O}(1/M)$ terms} 
In any perturbative approach, the dominant contributions are expected to arise from the LO corrections. Although, the
diagrams (a) and (b) (see Fig.~\ref{fig1}) formally give the leading chiral order contributions, in fact, they also 
contain the suppressed ${\mathcal O} (1/M)$ terms which are numerically commensurate with other NLO chiral order 
contributions to the cross-section. Likewise, the NLO chiral order diagrams (c) - (i) have similar kinematically
suppressed ${\mathcal O} (1/M)$ terms which effectively contribute as ${\mathcal O} (1/M^2)$, i.e., of NNLO chiral order. 
However, since the latter contribution is of higher order than our desired NLO accuracy, we ignore such terms in our results. 
It is important to realize that the anomalous proton's magnetic moment also contributes at NNLO accuracy in the TPE 
contribution to elastic cross-section (despite its appearances in the NLO Lagrangian) and therefore this magnetic moment is 
similarly neglected. In other words, in all our NLO expressions we may simply replace the outgoing lepton kinematical 
variables like $E^\prime$ and $\beta^\prime$ by the incoming variables $E$ and $\beta$. 

We first investigate only the diagrams (a) and (b). Their contribution to the radiative process is given by
\begin{widetext}
\begin{eqnarray}
		\delta^{(a)}_{\rm box}(Q^2) \!&=&\! \frac{2{\mathcal R}e\sum\limits_{spins}
        \bigg[\mathcal{M}^{(0)*}_{\gamma} {\mathcal M}^{(a)}_{\rm box}\bigg]}{\sum\limits_{spins}
        \left|\mathcal{M}^{(0)}_\gamma\right|^2}
		\nonumber\\
		&=&\! -\,4 \pi\alpha \left[\frac{Q^2}{Q^2+4E E^\prime}\right]  
        {\mathcal R}e\left\{\frac{1}{i}\int \frac{{\rm d}^4 k}{{(2 \pi)}^4}
        \frac{{\rm Tr}\left[(\slashed{p}+m_l)\,\not{\!v}\,(\slashed{p}^{\prime}+m_l)\,\slashed{v}\, 
        (\slashed{p}-\slashed{k}+m_l)\,\slashed{v}\right]}{(k^2+i0)\,
        [(Q-k)^2+i0]\,(k^2-2 k\cdot p+i0)\,(v\cdot k+i0)}\right\}\,,
        \label{a-SPA}
\end{eqnarray}
and
\begin{eqnarray} 
		\delta^{(b)}_{\rm xbox}(Q^2) \!&=&\! \frac{2{\mathcal R}e\sum\limits_{spins}
        \bigg[\mathcal{M}^{(0)*}_{\gamma} {\mathcal M}^{(b)}_{\rm xbox}\bigg]}{\sum\limits_{spins}
        \left|\mathcal{M}^{(0)}_\gamma\right|^2}
        \nonumber\\
		&=&\! -\,4 \pi\alpha \left[\frac{Q^2}{Q^2+4E E^\prime}\right]  
        {\mathcal R}e\left\{\frac{1}{i}\int \frac{{\rm d}^4 k}{{(2 \pi)}^4}
        \frac{{\rm Tr}\left[(\slashed{p}+m_l)\,\slashed{v}\,(\slashed{p}^{\, \prime}+m_l)\,\slashed{v}\, 
        (\slashed{p}+\slashed{k}-\slashed{Q}+m_l)\,\slashed{v}\right]}{(k^2+i0)\,
        [(Q-k)^2+i0]\,(k^2 + 2k\cdot p^\prime+i0)\,(v\cdot k+i0)}\right\}\,,
        \label{b-SPA}
\end{eqnarray}
\end{widetext}
respectively. Before proceeding to evaluate  the integrals, some comments about $\delta^{(a)}_{\rm box}$ and 
$\delta^{(b)}_{\rm xbox}$ corrections are warranted. Equations~\eqref{a-SPA} and \eqref{b-SPA} contain four-point
integrals function with two photon propagators. In the literature, the analytical evaluations of such types of 
four-point functions (albeit with all relativistic propagators) used SPA~\cite{Tsai:1961zz} to simplify 
calculations. Notably, two kinds of SPA have often been employed in the literature, namely, the one by Mo and 
Tsai~\cite{Tsai:1961zz,Mo:1968cg}, and the other by Maximon and Tjon~\cite{Maximon:1969nw,Maximon:2000hm}. Invoking 
such an approximation, while one of the exchanged photons is always considered ``soft", either with four-momenta 
$k=0$ or $k=Q$, and the other photon is considered ``hard", with $(k-Q)^2 \neq 0$ or $k^2 \neq 0 $, leads to two 
distinct kinematical domains of the TPE loop configurations. Note, however, the fact that the kinematical regions 
where both the photon exchanges are simultaneously soft or hard are ignored in the SPA methodology (see 
Ref.~\cite{Talukdar:2019dko} for a detailed discussion). SPA is an unwarranted approximation that could lead to 
unknown systematic uncertainties. On the one hand, SPA drastically simplifies the calculation of the four-point 
integrals without loss of information in the IR domain. On the other hand, SPA neglects finite contributions to the
TPE, which could prove to be a major drawback. Thus, when it comes to the precise estimation of the TPE effects, 
SPA results prove to be mostly unreliable. Therefore, in contrast to recent works based on similar perturbative 
evaluations of TPE loop diagrams, we do not take recourse to SPA for their exact analytical evaluation. Using IBP 
technique, the complete expressions of $\delta^{(a)}_{\rm box}$ and $\delta^{(b)}_{\rm xbox}$ can be expressed in 
terms of a string of three- and four-point integrals, defined by the generic master integral 
$I^{\pm}(p,\omega|n_1,n_2,n_3,n_4)$ (cf. Appendix~\ref{appA}), and are given by
\begin{widetext}
\begin{eqnarray}
		\delta^{(a)}_{\rm box}(Q^2) \!&=&\! 
        -\,8\pi\alpha\left[\frac{Q^2}{Q^2+4E E^\prime}\right] {\mathcal R}e\,
        \bigg\{E^\prime I^{-}(p,0|0,1,1,1)+E I^{-}(p,0|1,0,1,1) 
        - (E+E^\prime) I^-(p,0|1,1,0,1)
        \nonumber \\
		&& \hspace{3.9cm}  -\, (Q^2+8EE^\prime) I^{-}(p,0|1,1,1,0) 
  + (Q^2+8EE^\prime) E I^{-}(p,0|1,1,1,1)\bigg\}\,,\,\text{and}\qquad\,\,
		\label{delta-a}
\\
		\delta^{(b)}_{\rm xbox}(Q^2) \!&=&\! 
        -\,8\pi\alpha\left[\frac{Q^2}{Q^2+4E E^\prime}\right] {\mathcal R}e\, 
        \bigg\{E I^{+}(p^\prime,0|0,1,1,1) + E^\prime I^{+}(p^\prime,0|1,0,1,1) 
        - (E+E^\prime) I^+(p^\prime,0|1,1,0,1)
        \nonumber \\
	    && \hspace{3.9cm}  +\, (Q^2+8EE^\prime) I^{+}(p^\prime,0|1,1,1,0) 
        + (Q^2+8EE^\prime) E^\prime I^{+}(p^\prime,0|1,1,1,1) \bigg\}\,,\quad\,\,
     \label{delta-b} 
\end{eqnarray}
respectively. In deriving the above expressions, we have used the facts that $p \cdot Q=Q^2/2$ and 
$p^\prime\cdot Q=-Q^2/2$. Furthermore, as demonstrated in the Appendix~\ref{appB}, the four-point integrals 
$I^{-}(p,0|1,1,1,1)$ and $I^{+}(p^\prime,0|1,1,1,1)$ are decomposed {\it via} the method of partial fractions
into sums of simpler three-point ($I^\pm$) and four-point ($Z^\pm$) functions, namely,
\begin{eqnarray}
I^{-}(p,0|1,1,1,1)=\frac{1}{Q^2}\Big[I^{-}(p,0|1,0,1,1) + I^{-}(p,0|0,1,1,1) 
- 2Z^-(\Delta,i\sqrt{-Q^2}/2,m_l,E)\Big]\,, \quad \text{and}
\nonumber\\
I^{+}(p^\prime,0|1,1,1,1)=\frac{1}{Q^2}\Big[I^{+}(p^\prime,0|1,0,1,1) + I^{+}(p^\prime,0|0,1,1,1) 
- 2Z^+(\Delta^\prime,i\sqrt{-Q^2}/2,m_l,-E^\prime)\Big]\,,
\label{I1111}
\end{eqnarray}
\end{widetext}
where the four-vector $\Delta_\mu=-\Delta^\prime_\mu=(p-\frac{Q}{2})_\mu$. The explicit analytical  expressions 
of the $I^\pm$ and $Z^\pm$ integral functions appearing above are rather elaborate and, therefore, relegated to 
Appendix~\ref{appB}.

\subsection{Leading chiral order TPE corrections, i.e., ${\mathcal O}(\alpha)$ } 
\label{LO-Contri}
First, we extract the {\it true} LO terms from the (a) and (b) diagrams to determine the leading finite TPE 
contributions. For this purpose we eliminate all ${\mathcal O}(1/M)$ terms by substituting $E^\prime=E$ and 
$\beta^\prime=\beta$, which yields the true LO sum of the TPE diagrams in HB$\chi$PT:
\begin{eqnarray}
\delta^{\rm (0)}_{\gamma\gamma}(Q^2) \!\!&=&\!\!  \left[\delta^{(a)}_{\rm box}(Q^2) + \delta^{(b)}_{\rm xbox}(Q^2) \right]_{\text{LO}}
\end{eqnarray} 
\begin{eqnarray}
&=&\! 32\pi\alpha E \left[\frac{Q^2}{Q^2+4E^2}\right] {\mathcal R}e \left[ I(Q|1,1,0,1) \right]_{\rm LO}\,.\quad\,\,\,
\label{ab-LO}
\end{eqnarray} 
Here we note that the integral, $I(Q|1,1,0,1) \equiv I^-(p,0|1,1,0,1) \equiv I^+(p^\prime,0|1,1,0,1)$,  
Eq.~\eqref{I(1101)},  is solely a function of the squared momentum transfer $Q^2$. 
Furthermore, we use the facts that at LO the real parts of the $I^-$ and $Z^-$ integrals appearing in the 
expression for $\delta^{(a)}_{\rm box}$, Eq.~\eqref{delta-a}, completely cancel with those from the integrals 
$I^+$ and $Z^+$, respectively, appearing in $\delta^{(b)}_{\rm xbox}$, Eq.~\eqref{delta-b}. This is evident 
from our explicit expressions for these integrals displayed in  Eqs.~\eqref{IP-1011} - \eqref{deltaZ+}. Thus, 
the only surviving term in the LO TPE contribution corresponds to the non-vanishing three-point function 
$I(Q|1,1,0,1)$ given by 
\begin{equation}
I(Q|1,1,0,1)=-\frac{1}{16} \sqrt{\frac{1}{-Q^2}} + {\mathcal O}\left(\frac{1}{M^2}\right)\,.
\end{equation}
This means that the true LO chiral result boils down to 
\begin{eqnarray}
\delta^{\rm (0)}_{\gamma\gamma}(Q^2) = \pi \alpha \frac{\sqrt{{-Q^2}}}{2E}\left[\frac{1}{1+\frac{Q^2}{4 E^2}}\right]\,,
\label{eq:ab-LO}
\end{eqnarray}
which bears a close resemblance to the well-known McKinley-Feshbach contribution~\cite{McKinley:1948zz} 
for relativistic electrons when $\beta \simeq 1$, given by
\begin{eqnarray}
\delta_F (Q^2) \!&=&\! \pi \alpha \beta  \sin{\frac{\theta_{lab}}{2}} 
\left[\frac{1-\sin{ \frac{\theta_{lab}}{2}}}{1-\beta^2 \sin^2{\frac{\theta_{lab}}{2}}}\right] 
\nonumber \\
&=&\! \pi\alpha \frac{\sqrt{-Q^2}}{2E} \left[\frac{1 - \frac{\sqrt{-Q^2}}{2\beta E}}{1+\frac{Q^2}{4E^2}} \right]\,,
\label{Feshbach}
\end{eqnarray}
where $\theta_{lab}$ refers to the lepton scattering angle in the {\it lab}-frame. The above 
result~\cite{McKinley:1948zz} was originally obtained in the context of potential scattering in non-relativistic 
quantum mechanics employing second-order Born approximation. Given the fact that the SPA approach in HB$\chi$PT 
pursued in Ref.~\cite{Talukdar:2019dko} had completely missed out on the McKiney Feshbach contribution, our exact LO TPE result is already a significant improvement over the SPA results. 
Furthermore, an important feature of our HB$\chi$PT is the absence of IR divergence at true LO:
\begin{eqnarray}
	\delta^{(0)}_{\gamma\gamma}(Q^2)\Big|_{\rm IR} \!\!&=&\!\! -\,16 \pi \alpha E\, {\mathcal R}e \, \big[ I^{-}(p,0|1,0,1,1) 
	\\
	&&\hspace{2cm} +\, I^{+}(p^\prime,0|1,0,1,1)\big]_{\rm LO} = 0\,.
	\nonumber
\end{eqnarray}
Finally, we note that Ref.~\cite{Peset:2021iul} in their Eq.~(4.13)  obtained a McKinley Feshbach result for  
$\beta \ll 1$, derived   
from the electron-muon scattering TPE result of Ref.~\cite{Kaiser:2010zz}.     
This, however, contrasts 
with the original McKinley Feshbach
derivation which assumed that $\beta \simeq 1$. 

\begin{figure*}
\begin{center}
		\includegraphics[scale=0.58]{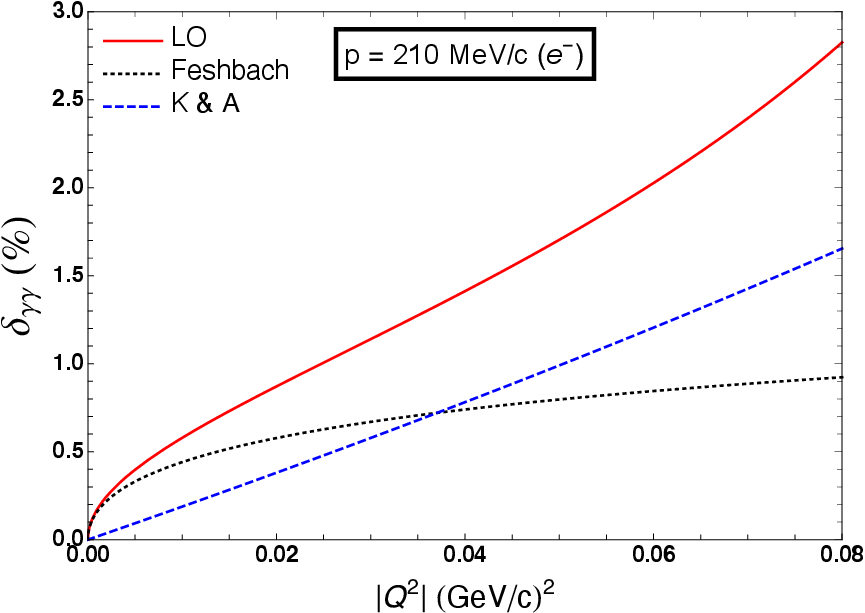}\qquad
		\includegraphics[scale=0.58]{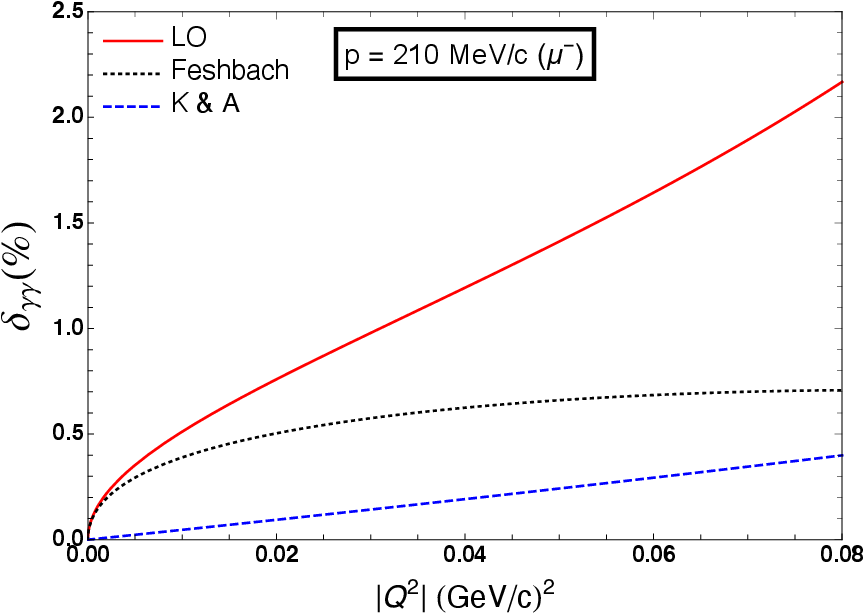} 

        \vspace{1.5cm}
  
        \includegraphics[scale=0.58]{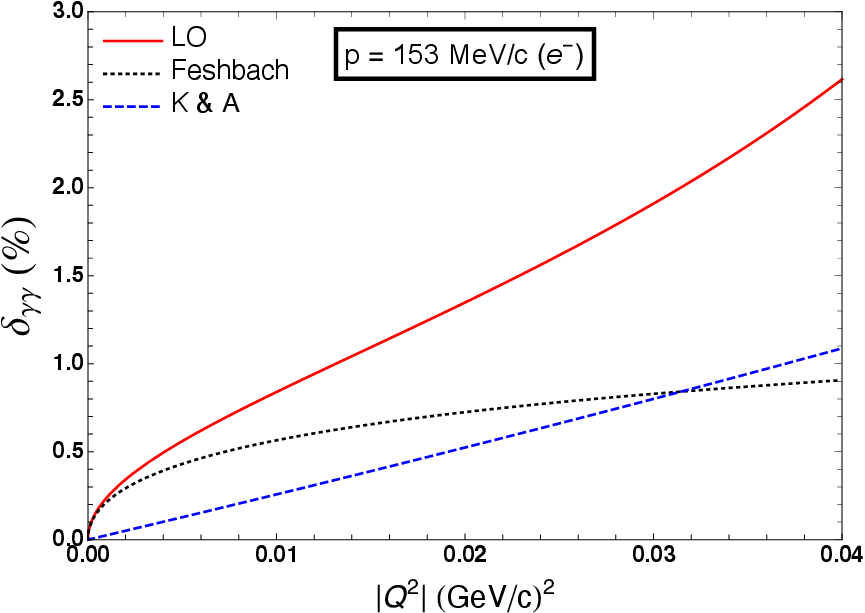}\qquad
		\includegraphics[scale=0.58]{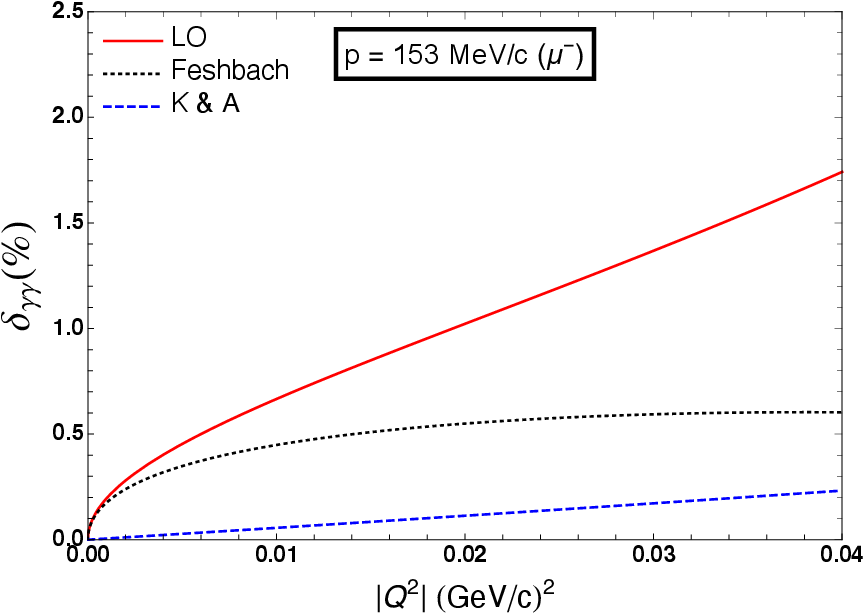}

        \vspace{1.5cm}
  
        \includegraphics[scale=0.58]{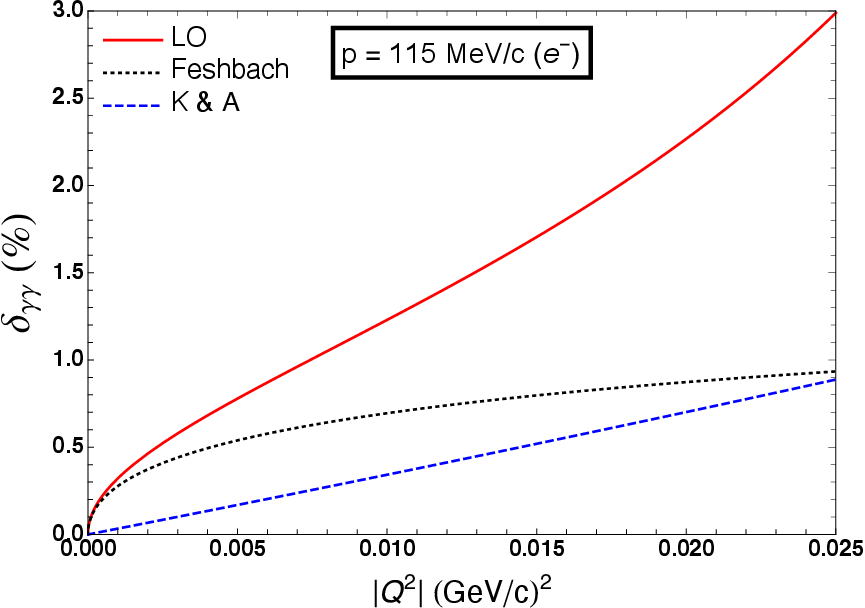}\qquad
		\includegraphics[scale=0.58]{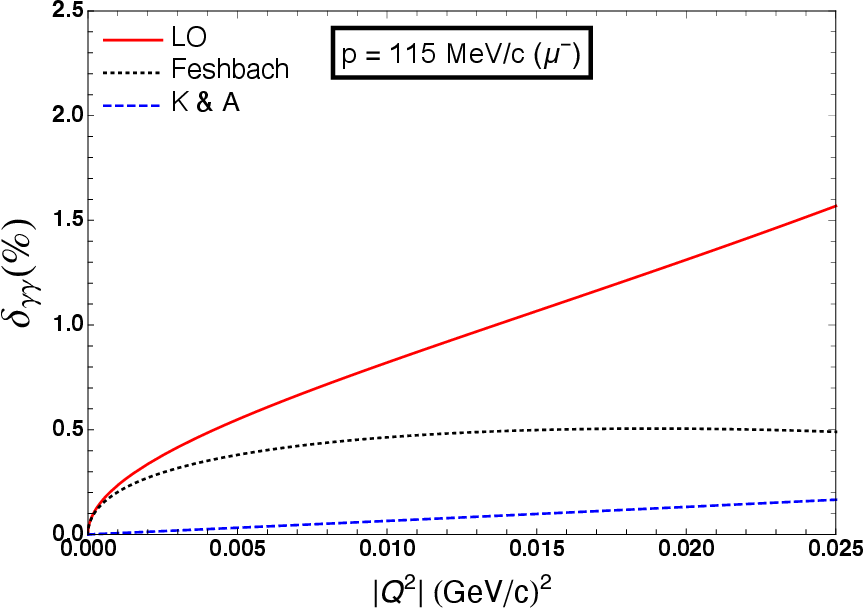} 
		\caption{The fractional LO TPE contributions (expressed in percentage) to the elastic lepton-proton scattering 
                 cross-section [i.e., of ${\mathcal O}(\alpha)$] in HB$\chi$PT, Eq.~\eqref{eq:ab-LO}. The left (right) 
                 panel displays the results for $e$-p ($\mu$-p) scattering versus the squared four-momentum transfer 
                 $|Q^2|$, for the MUSE beam momenta of $|\vec p\,|=p=115,\, 153,\, 210$~MeV/c. Each plot covers the 
                 full kinematical scattering range, $0<|Q^2|<|Q^2_{\rm max}|$ when $\theta\in[0,\pi]$. For comparison, 
                 the results for the well-known McKinley-Feshbach contribution, Eq.~\eqref{Feshbach}, (denoted as 
                 ``Feshbach"), and that of Ref.~\cite{Koshchii:2017dzr} (denoted as ``K \& A"), are also displayed. }
  \label{LO-results}
\end{center}
\end{figure*}

Fig.~\ref{LO-results} displays the numerical results for the true LO TPE fractional corrections,\footnote{We 
re-emphasize that in the HB$\chi$PT framework, only the true LO [i.e., of ${\mathcal O}(1/M^0)$]  corrections 
are regarded as the LO chiral contributions. All other ${\mathcal O}(1/M)$ corrections are treated as NLO 
contributions.} Eq.~\eqref{eq:ab-LO} (with respect to the Born contribution), for specific MUSE choices of the 
incoming lepton beam momenta, both for e-p and $\mu$-p scattering. Our displayed results cover the full 
kinematical scattering range $0<|Q^2|<|Q^2_{\rm max}|$, where~\cite{Talukdar:2019dko} 
$$|Q^2_{\rm max}|=\frac{4M^2\beta^2E^2}{m^2_l+M^2+2ME}\,.$$ Evidently, the TPE corrections for muon and electron 
are different, as they depend on the non-zero lepton mass $m_l$.

\vspace{-0.2cm}

\subsection{Next-to-leading order TPE corrections, i.e., ${\mathcal O}(\alpha/M)$}
The NLO TPE fractional radiative corrections to elastic cross-section are of ${\mathcal O}(\alpha/M)$ contribution from
each TPE diagram. As discussed earlier, they include both the ${\mathcal O}(1/M)$ parts of the TPE diagrams, (a) and (b),
as well as the NLO chiral order TPE diagrams, (c) - (i), either with one proton-photon vertex stemming from the NLO 
Lagrangian ${\mathcal L}^{(1)}_{\pi N}$, Eq.~\eqref{eq:Lchiral1}, or with an insertion of an NLO proton propagator. 

First, we display our result for the ${\mathcal O}(1/M)$ parts of the (a) plus (b) diagrams. This part is 
obtained by eliminating 
the true LO contributions $\delta^{\rm (0)}_{\gamma\gamma}$ from the complete contribution from (a) and  (b) diagrams, 
$\delta^{\rm (ab)}_{\gamma\gamma}$. It is  represented by the following expression:
\begin{widetext}
\begin{eqnarray}
\delta^{\rm (ab;\, 1/M)}_{\gamma\gamma}(Q^2) \!& = &\! \delta^{\rm (ab)}_{\gamma\gamma}(Q^2) 
- \delta^{\rm (0)}_{\gamma\gamma}(Q^2) = \delta^{(a)}_{\rm box}(Q^2) + \delta^{(b)}_{\rm xbox}(Q^2)
-\left[ \delta^{(a)}_{\rm box}(Q^2) + \delta^{(b)}_{\rm xbox}(Q^2)\right]_{\text{LO}}
\nonumber\\
&=&\! -\, 16 \pi \alpha E\, {\mathcal R}e \left[\delta^{(1/M)}I^{+}(p^\prime,0|1,0,1,1)\right] 
+ 8 \pi \alpha \frac{Q^2}{M}\, {\mathcal R}e\, \left[I^{(0)}(p,0|1,0,1,1)\right] 
\nonumber\\
&&\! -\, 16 \pi \alpha E\,{\mathcal R}e\,\left[\delta^{(1/M)}I^{-}(p,0|0,1,1,1)  
+ \delta^{(1/M)}I^{+}(p^\prime,0|0,1,1,1)\right] 
\nonumber \\
&&\! +\, 4 \pi \alpha \frac{Q^2}{M}\left[\frac{8E^2}{Q^2+4E^2}\right]\, {\mathcal R}e\, \left[I^{(0)}(p,0|0,1,1,1)\right] 
+ 8 \pi \alpha \frac{Q^2}{M} \!\left[\frac{Q^2}{Q^2+4E^2}\right] \! \left[\frac{Q^2-4 E^2} {Q^2+4E^2}\right]
\!{\mathcal R}e\left[ I(Q|1,1,0,1) \right]
\nonumber\\
&&\! -\, 8 \pi \alpha \frac{Q^2}{M}\left[\frac{Q^2+8 E^2}{Q^2+4E^2}\right]\,
{\mathcal R}e\, \left[ Z^{(0)}(\Delta,i\sqrt{-Q^2}/2,m_l,E)\right] 
\nonumber \\
&&\! +\, 16\pi\alpha E\left[\frac{Q^2+8 E^2}{Q^2+4E^2}\right]\,{\mathcal R}e\, 
\left[ \delta^{(1/M)} Z^{-}(\Delta,i\sqrt{-Q^2}/2,m_l,E) 
+ \delta^{(1/M)} Z^{+}(\Delta^\prime,i\sqrt{-Q^2}/2,m_l,-E^\prime)\right]
\nonumber\\
&&\! +\, {\mathcal O}\left(M^{-2}\right)\,.
\label{delta-ab2}
\end{eqnarray} 

\vspace{-0.3cm}

\end{widetext}
Here, the functions $I^{(0)}$ and $Z^{(0)}$ [cf. Eqs.~\eqref{H7} and \eqref{Z0}] denote the LO parts of the three-, 
and four-point functions $I^-$ and $Z^-$, respectively, as displayed in Appendix~\ref{appB}. Also, 
$\delta^{(1/M)}I^{\pm}$ [cf. Eqs.~\eqref{eq:delta_I-0111} and \eqref{eq:delta_I+0111}] and $\delta^{(1/M)}Z^{\pm}$ 
[cf. Eqs.~\eqref{deltaZ-} and \eqref{deltaZ+}], denote the ${\mathcal O}(1/M)$ parts of the functions $I^\pm$ and 
$Z^\pm$, respectively. It is noteworthy that Eq.~\eqref{delta-ab2} is IR-singular due to the presence of the 
IR-divergent integrals $I^{-}(p,0|1,0,1,1)\equiv I^{(0)}(p,0|1,0,1,1)$ and $I^{+}(p^\prime,0|1,0,1,1)$ [cf. 
Eqs.~\eqref{IP-1011} and \eqref{IP+1011_a}]. The LO IR-divergent terms arising for these integrals, however, cancel
each other leaving only ${\mathcal O}(1/M)$ residual IR divergences arising from terms containing the integral 
$I^{+}(p^\prime,0|1,0,1,1)$. Thus, the resulting IR divergence from Eq.~\eqref{delta-ab2}, extracted using 
dimensional regularization in $D=4-2\epsilon$ dimension, with pole $\epsilon<0$ and arbitrary renormalization scale 
$\mu$, has the form:
\begin{widetext}
\begin{eqnarray} 
 \delta^{\rm (box)}_{\rm IR}(Q^2) \!& \equiv &\!
	\delta^{\rm (ab;\, 1/M)}_{\gamma\gamma}(Q^2)\Big|_{\rm IR}    = 
	-\, 16 \pi \alpha E\, {\mathcal R}e \left[\delta^{(1/M)}I^{+}(p^\prime,0|1,0,1,1)\right]_{\rm IR} 
	+ 8 \pi \alpha \frac{Q^2}{M}\, {\mathcal R}e\, \left[I^{(0)}(p,0|1,0,1,1)\right]_{\rm IR} 
	\nonumber\\
	&=&\! -\,\frac{\alpha Q^2}{2\pi M E \beta^2} \left[\frac{1}{\epsilon}-\gamma_E+\ln\left(\frac{4\pi\mu^2}{m_l^2}\right) \right] 
	\left\{1+ \left(\beta-\frac{1}{\beta}\right)\ln\sqrt{\frac{1+\beta}{1-\beta}}\,\right\} 
	+ {\mathcal O}\left(\frac{1}{M^2}\right)\,.
	\label{deltaIRbox}
\end{eqnarray}  
\end{widetext}
where $\gamma_E=0.577216...$  is the Euler-Mascheroni constant. The above NLO TPE contributions constitute the only
IR-divergent terms $\delta^{\rm (box)}_{\rm IR}$ that arise in our calculations. These IR divergences exactly cancel 
with the soft-bremsstrahlung IR-divergent counterparts $\delta^{\rm (soft)}_{\rm IR}$ originating from the 
${\mathcal O}(1/M)$ kinematical part of the interference contributions of the LO soft-bremsstrahlung radiation 
diagrams, such that, $\delta^{\rm (soft)}_{\rm IR}=-\delta^{\rm (box)}_{\rm IR}$. An explicit demonstration of this 
cancellation will be detailed in a future publication. We remark that in the IR divergent expression, 
Eq.~(\ref{deltaIRbox}), we include a constant term proportional to $\ln \left(\mu^2/m_l^2\right) $ whereas in 
Refs.~\cite{Talukdar:2019dko,Talukdar:2020aui,Cao:2021nhm} the corresponding IR divergence was written with a 
$Q^2$-dependent term proportional to $\ln \left(-Q^2/\mu^2\right) $. The difference produces a term proportional to 
$\ln \left(-Q^2/m_l^2 \right) $ which affects the finite part of the TPE contribution markedly. 

\begin{figure*}
\begin{center}
		\includegraphics[scale=0.58]{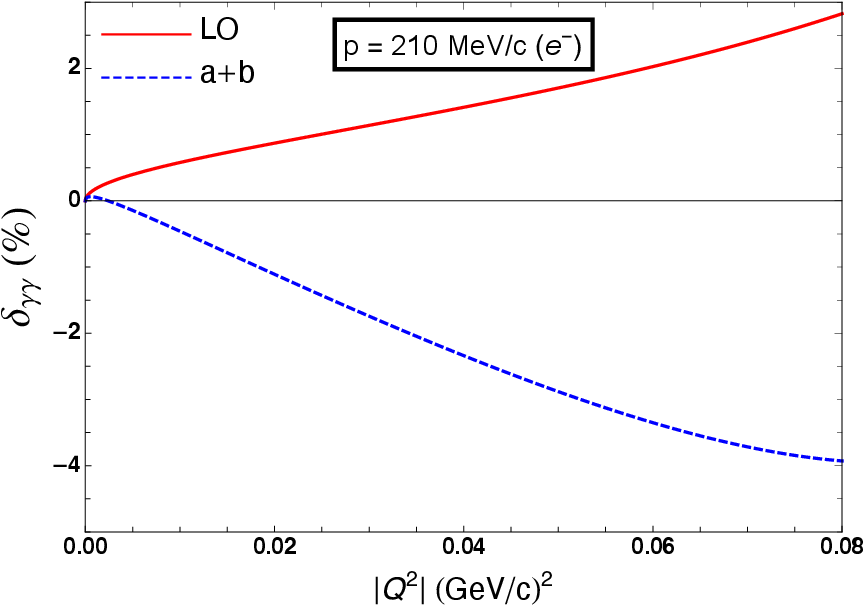}\qquad
		\includegraphics[scale=0.58]{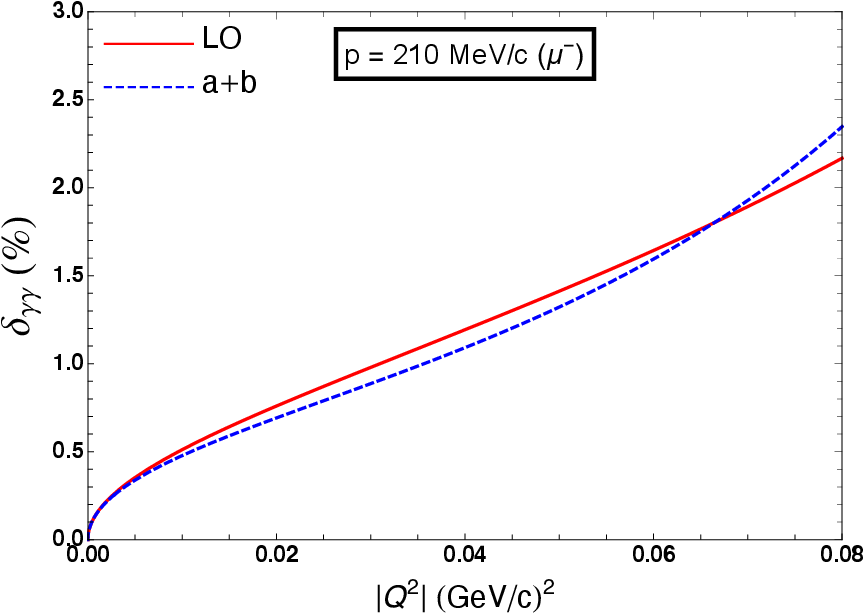} 
  
        \vspace{1.5cm}
        
        \includegraphics[scale=0.58]{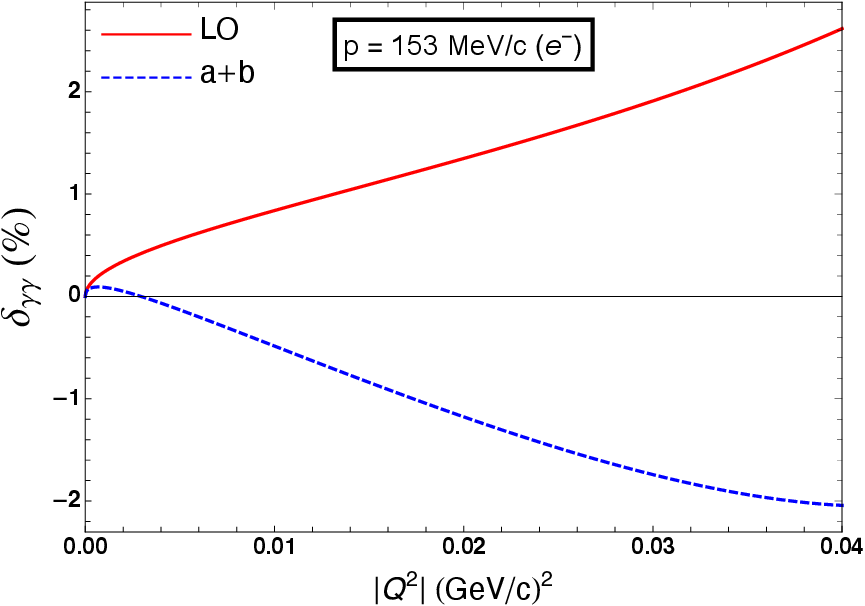}\qquad
		\includegraphics[scale=0.58]{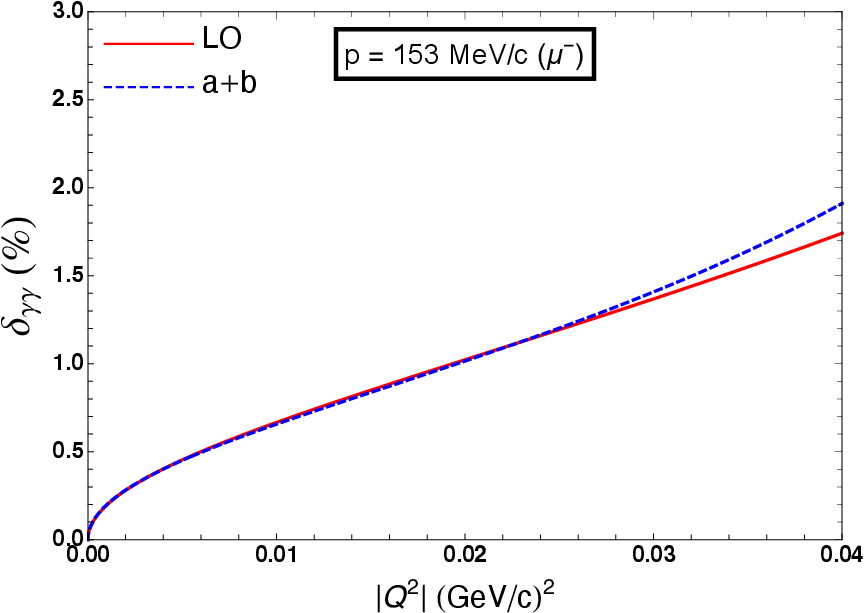} 

       \vspace{1.5cm}
       
        \includegraphics[scale=0.58]{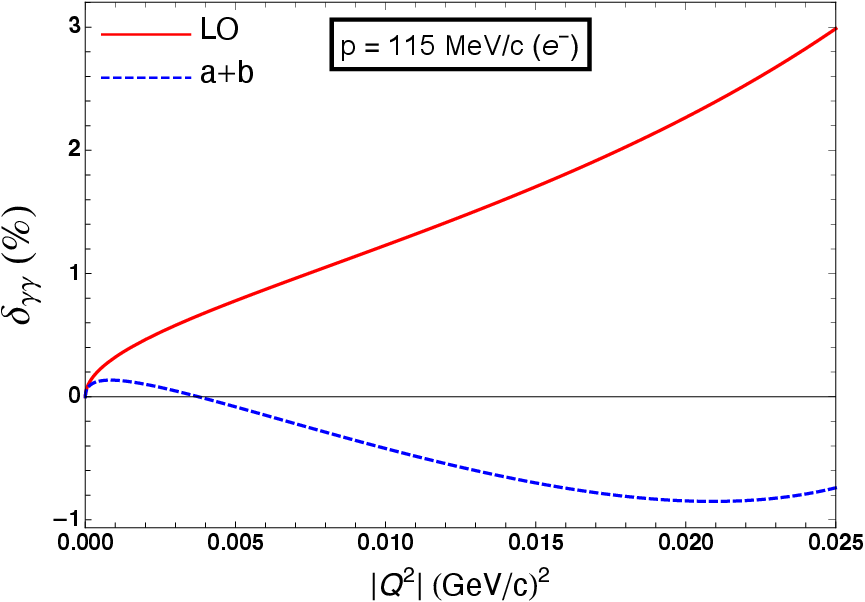}\qquad
		\includegraphics[scale=0.58]{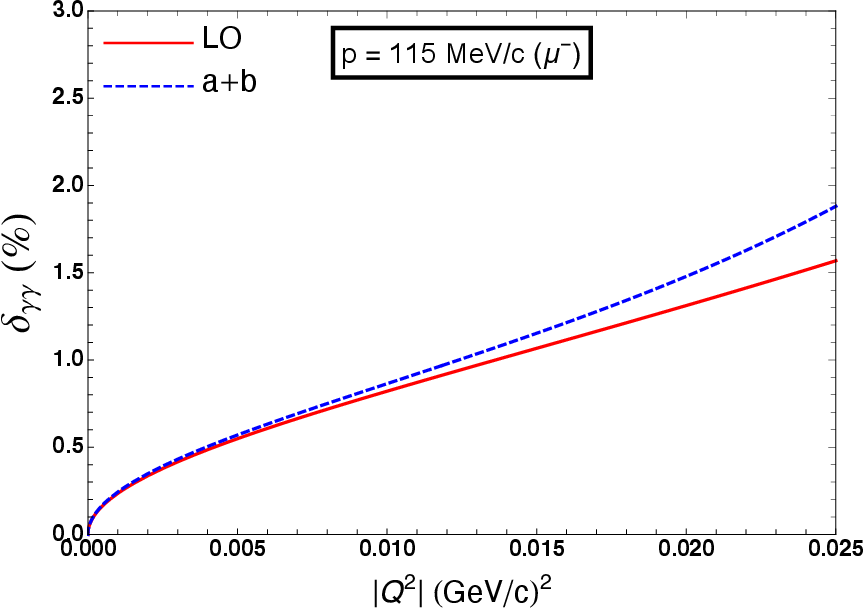} 
		\caption{Comparison between the true LO TPE corrections $\delta^{(0)}_{\gamma\gamma}$  (denoted above as ``LO") 
                 and the full TPE corrections from the (a) and (b) diagrams $\delta^{(a+b)}_{\gamma\gamma}$ (denoted 
                 above as ``a+b") in HB$\chi$PT. The left (right) panel displays the results for e-p  ($\mu$-p) 
                 scattering versus the squared four-momentum transfer $|Q^2|$, for the MUSE beam momenta of 
                 $|\vec p\,|=p=115,\, 153,\, 210$~MeV/c. Each plot covers the full kinematical scattering range, 
                 $0<|Q^2|<|Q^2_{\rm max}|$  with $\theta\in[0,\pi]$. }
\label{LO-LO1/M-results} 
\end{center}
\end{figure*}

In Fig.~\ref{LO-LO1/M-results}, we present a comparison between the true LO TPE corrections 
$\delta^{(0)}_{\gamma\gamma}$ and the full TPE corrections arising from the (a) and (b) diagrams 
$\delta^{(a+b)}_{\gamma\gamma}$ which also include the proton propagator ${\mathcal O}(1/M)$ NLO contributions. As 
observed in this figure the NLO part of diagrams (a) and (b) in the electron-proton scattering case is negative and 
dominates over the positive true LO part. In contrast, for the muon-proton scattering the NLO parts of these diagrams 
are quite small and barely alter the LO results. 

\begin{figure*}[tbp]
\begin{center}
		\includegraphics[scale=0.58]{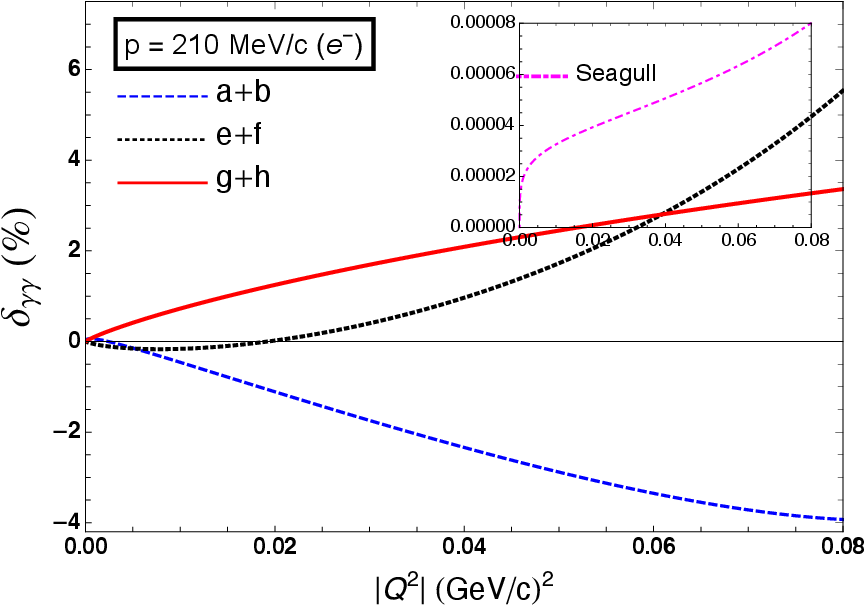}\qquad
		\includegraphics[scale=0.58]{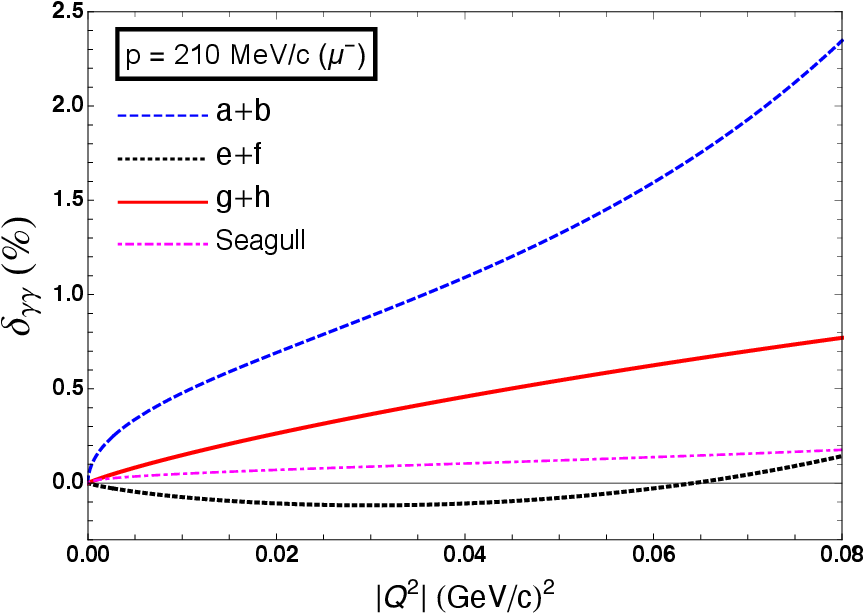} 
  
        \vspace{1.5cm}
        
        \includegraphics[scale=0.58]{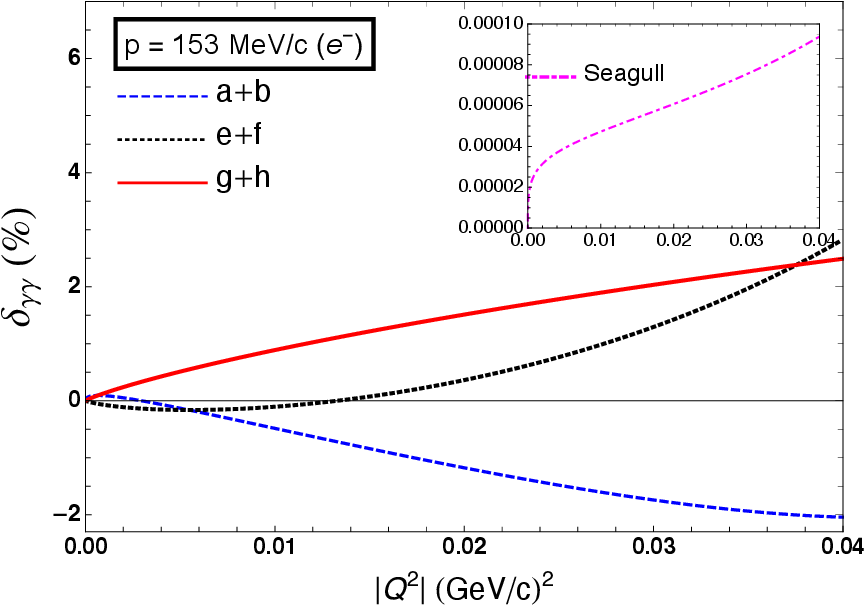}\qquad
		\includegraphics[scale=0.58]{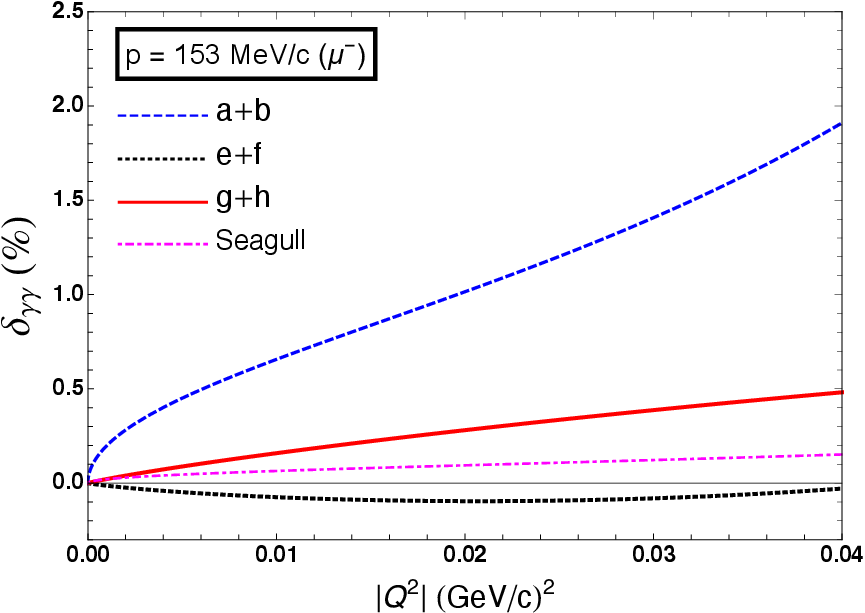} 

       \vspace{1.5cm}
       
        \includegraphics[scale=0.58]{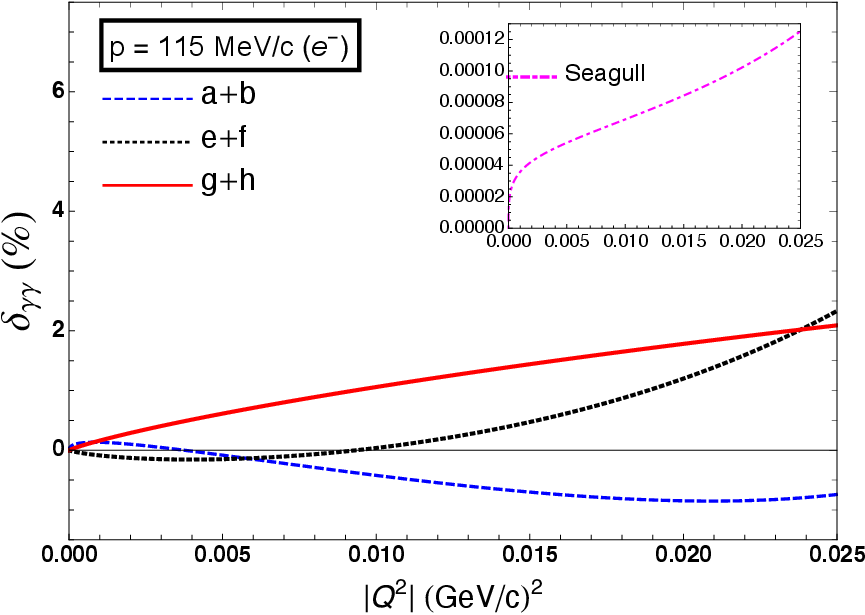}\qquad
		\includegraphics[scale=0.58]{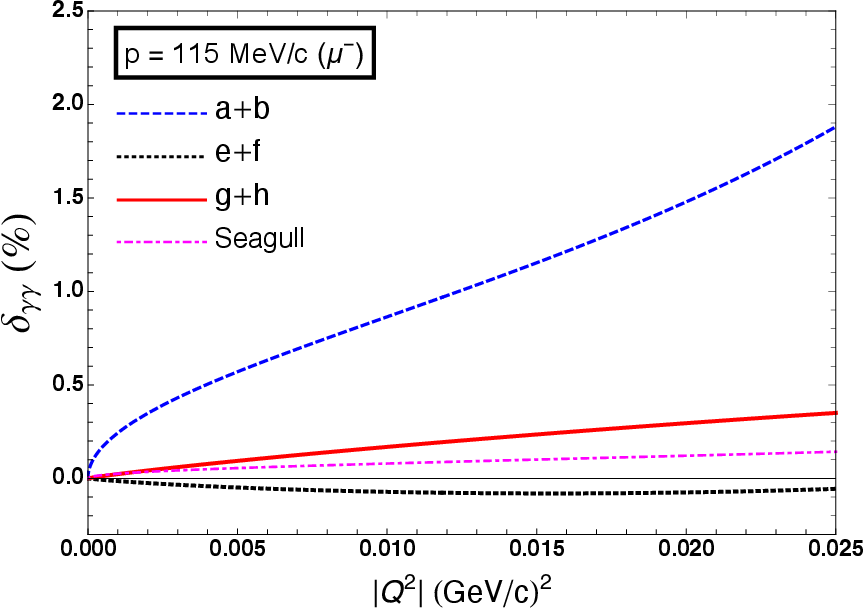} 
		\caption{Various diagrammatic contributions to the TPE radiative corrections to the elastic cross-section. 
                 The pair-wise contributions from the TPE diagrams, namely, (a) + (b), (e) + (f), and (g) + (h) are 
                 displayed, along with that of the seagull diagram (i). The left (right) panel displays the results 
                 for e-p  ($\mu$-p) scattering versus the squared four-momentum transfer $|Q^2|$, for the MUSE beam 
                 momenta of $|\vec p\,|=p=115,\, 153,\, 210$~MeV/c. Each plot covers the full kinematical scattering 
                 range, $0<|Q^2|<|Q^2_{\rm max}|$  with $\theta\in[0,\pi]$. The seagull contribution for the electron
                 scattering is tiny and is displayed in the inset plots.}
\label{LO-NLO-results-1} 
\end{center}
\end{figure*}

Next, we turn to the NLO chiral order diagrams (c) and (d), as well as (e) and (f), obtained by replacing one LO vertex  
with one NLO vertex, namely,
\begin{widetext}
\begin{eqnarray}
\delta^{(c)}_{\rm box}(Q^2) \!&=&\! \frac{2{\mathcal R}e\sum\limits_{spins}
\bigg[\mathcal{M}^{(0)*}_{\gamma} \widetilde{\mathcal M}^{(c)}_{\rm box}\bigg]}{\sum\limits_{spins}
\left|\mathcal{M}^{(0)}_\gamma\right|^2} 
\nonumber \\
&=&\! -\,\frac{2 \pi\alpha}{M}\left[\frac{Q^2}{Q^2+4EE^\prime}\right]  
{\mathcal R}e\, \left\{\frac{1}{i}\int \frac{{\rm d}^4 k}{{(2 \pi)}^4}
\frac{{\rm Tr}\left[(\slashed{p}+m_l)\,\slashed{v}\,(\slashed{p}^{\, \prime}+m_l)\,\slashed{v}\, 
(\slashed{p}-\slashed{k}+m_l)\, \slashed{k}\right]}{(k^2+i0)\,
\left[(Q-k)^2+i0\right]\,(k^2-2 k\cdot p+i0)\,(v\cdot k+i0)}\right\}\,,
\label{delta-c1}
\\
\delta^{(d)}_{\rm xbox}(Q^2) \!&=&\! \frac{2{\mathcal R}e\sum\limits_{spins}
\bigg[\mathcal{M}^{(0)*}_{\gamma} \widetilde{\mathcal M}^{(d)}_{\rm xbox}\bigg]}{\sum\limits_{spins}
\left|\mathcal{M}^{(0)}_\gamma\right|^2}
\nonumber \\
&=&\! -\, \frac{2 \pi\alpha}{M}\left[\frac{Q^2}{Q^2+4EE^\prime}\right]  
{\mathcal R}e\, \left\{\frac{1}{i}\int \frac{{\rm d}^4 k}{{(2 \pi)}^4}\frac{{\rm Tr}\left[(\slashed{p}+m_l)\,\slashed{v}\,
(\slashed{p}^{\, \prime}+m_l)\,\slashed{k}\,(\slashed{p}+\slashed{k}-\slashed{Q}+m_l)\,\slashed{v}\right]}{(k^2+i0)\,
\left[(Q-k)^2+i0\right]\,(k^2+2 k\cdot p^{\prime}+i0)\,(v\cdot k+i0)}\right\}\,,
\label{delta-d1}
\\
\delta^{(e)}_{\rm box}(Q^2) \!&=&\! \frac{2{\mathcal R}e\sum\limits_{spins}
\bigg[\mathcal{M}^{(0)*}_{\gamma} \widetilde{\mathcal M}^{(e)}_{\rm box}\bigg]}{\sum\limits_{spins}
\left|\mathcal{M}^{(0)}_\gamma\right|^2}
\nonumber\\
&=&\! -\, \frac{2 \pi\alpha}{M}\left[\frac{Q^2}{Q^2+4EE^\prime}\right]  
{\mathcal R}e\, \left\{\frac{1}{i}\int \frac{{\rm d}^4 k}{{(2 \pi)}^4}\frac{{\rm Tr}\left[(\slashed{p}+m_l)\,\slashed{v}\,
(\slashed{p}^{\, \prime}+m_l)\,(\slashed{k} +\slashed{Q})\, (\slashed{p}-\slashed{k}+m_l)\, \slashed{v}\right]}{(k^2+i0)\,
\left[(Q-k)^2+i0\right]\,(k^2-2 k\cdot p+i0)\,(v\cdot k+i0)}\right\}\,,
\label{delta-E1}
\end{eqnarray}
and
\begin{eqnarray}
\delta^{(f)}_{\rm xbox}(Q^2) \!&=&\! \frac{2{\mathcal R}e\sum\limits_{spins}
\bigg[\mathcal{M}^{(0)*}_{\gamma} \widetilde{\mathcal M}^{(f)}_{\rm xbox}\bigg]}{\sum\limits_{spins}
\left|\mathcal{M}^{(0)}_\gamma\right|^2}
\nonumber\\
&=&\! -\,\frac{2 \pi\alpha}{M}\left[\frac{Q^2}{Q^2+4EE^\prime}\right]  
{\mathcal R}e\, \left\{\frac{1}{i}\int \frac{{\rm d}^4 k}{{(2 \pi)}^4}\frac{{\rm Tr}\left[(\slashed{p}+m_l)\,\slashed{v}\,
(\slashed{p}^{\, \prime}+m_l)\,\slashed{v}\, (\slashed{p}+\slashed{k}-\slashed{Q}+m_l)\, 
(\slashed{k}+\slashed{Q})\right]}{(k^2+i0)\,\left[(Q-k)^2+i0\right]\,(k^2+2 k\cdot p^\prime+i0)\,(v\cdot k+i0)}\right\}\,.
\label{delta-f1}
\end{eqnarray}
\end{widetext}
These TPE diagrams also have higher-order contributions that we neglect in our NLO evaluations. Using IBP methods the 
contributions from the (c) and (d) diagrams can be expressed in terms of the real part of the IR-finite three-point 
integral $I(Q|1,1,0,1)$ given in Eq.(27) [also see Eq.~\eqref{I(1101)}]:

\vspace{-0.6cm}

\begin{eqnarray}
\delta^{(c)}_{\rm box}(Q^2) \!&=&\! 4\pi\alpha \frac{Q^2}{M}\, {\mathcal R}e\, \left[I(Q|1,1,0,1)\right]\,, \quad \text{and}
\label{delta-c}
\\
\delta^{(d)}_{\rm xbox}(Q^2) \!&=&\! -\, 4\pi\alpha \frac{Q^2}{M}\,{\mathcal R}e\, \left[I(Q|1,1,0,1)\right]\,,
\label{delta-d}
\end{eqnarray}	
which means that their sum is 
\begin{equation}
\delta^{(c+d)}_{\gamma\gamma}(Q^2) = \delta^{(c)}_{\rm box}(Q^2) + \delta^{(d)}_{\rm xbox}(Q^2) = 0\,.
\label{delta_c+d}
\end{equation}
Thus, the net contribution from the (c) and (d) diagrams to the TPE radiative corrections vanishes. The other two 
diagrams, namely, (e) and (f) diagrams are given by the following expressions:
\begin{widetext}
\begin{eqnarray}
\delta^{(e)}_{\rm box}(Q^2) \!&=&\!
-\,\frac{4\pi\alpha}{M}\left[\frac{Q^2}{Q^2+4EE^\prime}\right] {\mathcal R}e\,
\bigg\{\!-2(Q^2+2E^2)\, I^{-}(p,0|0,1,1,1) + 4 E^2\, I^{-}(p,0|1,0,1,1) + (Q^2-4 E^2)
\nonumber \\
&&\hspace{4cm} \times\, I(Q|1,1,0,1) + 4 E Q^2\, I^{-}(p,0|1,1,1,0) - 4 E^2 Q^2\, I^{-}(p,0|1,1,1,1)\bigg\}
\nonumber \\ 
&=&\! -\,\frac{4\pi\alpha}{M}\left[\frac{Q^2}{Q^2+4E^2}\right] {\mathcal R}e\,
\bigg\{\!-2(Q^2+4E^2)\, I^{(0)}(p,0|0,1,1,1) + (Q^2-4 E^2)\, I(Q|1,1,0,1)
\nonumber \\
&&\hspace{3.9cm}  +\, 4 E Q^2\, I(Q|1,1,1,0) + 8 E^2\, Z^{(0)}(\Delta,i\sqrt{-Q^2}/2,m_l,E) \bigg\}
+ {\mathcal O}\left(\frac{1}{M^2}\right) \,, \quad\,\,
\label{delta-E}
\end{eqnarray}
and
\begin{eqnarray}
\delta^{(f)}_{\rm xbox}(Q^2) \!&=&\!
-\, \frac{4\pi\alpha}{M}\left[\frac{Q^2}{Q^2+4EE^\prime}\right] {\mathcal R}e\, 
\bigg\{2(Q^2+2 E^2)\, I^{+}(p^\prime,0|0,1,1,1) - 4 E^2\, I^{+}(p^\prime,0|1,0,1,1) - (Q^2-4 E^2)
\nonumber \\ 
&&\hspace{4cm} \times\, I(Q|1,1,0,1) + 4 E Q^2\, I^{+}(p^\prime,0|1,1,1,0) + 4 E^2 Q^2\, 
I^{+}(p^\prime,0|1,1,1,1)\bigg\} 
\nonumber \\
&=&\! -\, \frac{4\pi\alpha}{M}\left[\frac{Q^2}{Q^2+4E^2}\right] {\mathcal R}e\,
\bigg\{\!-2(Q^2+4 E^2)\, I^{(0)}(p,0|0,1,1,1)  - (Q^2-4 E^2)\, I(Q|1,1,0,1) 
\nonumber \\
&&\hspace{3.9cm}  +\, 4 E Q^2\, I(Q|1,1,1,0) + 8 E^2\, Z^{(0)}(\Delta,i\sqrt{-Q^2}/2,m_l,E) \bigg\}
+ {\mathcal O}\left(\frac{1}{M^2}\right)\,.\quad\,\,
\label{delta-f}
\end{eqnarray}
\end{widetext}
The IR-finite three-point integrals, $I(Q|1,1,1,0)\equiv I^{-}(p,0|1,1,1,0) \equiv I^{+}(p^\prime,0|1,1,1,0)$, 
are purely relativistic and depend only on the momentum transfer $Q^2$ [cf. Eq.~(\ref{I1110})]. It is important to 
note that the IR-divergent functions, namely, $I^-(p,0|1,0,1,1)$ and $I^+(p^\prime,0|1,0,1,1)$ [arising due to the 
decompositions, Eq.~\eqref{I1111}], drop out from the above expressions leading to IR-finite contributions from each
of (e) and (f) NLO diagrams. Thus, the net contribution to the TPE radiative corrections from the (e) and (f) 
diagrams is finite and displayed in Fig.~\ref{LO-NLO-results-1}. Notably, the sizable corrections in the case of 
electron-proton scattering arise primarily due to the dominance of the integral $I(Q|1,1,1,0)$ at large $Q^2$ values.

Next, we display our analytical results for the (g) and (h) NLO diagrams containing the NLO proton propagator 
insertions, where we employ Eqs.~(\ref{eq:I1-T1_tranform}) and (\ref{eq:T1_tensors}), 
namely,
\begin{widetext}
\begin{eqnarray}
\delta^{(g)}_{\rm box}(Q^2) \!&=&\! \frac{2{\mathcal R}e\sum\limits_{spins}
\bigg[\mathcal{M}^{(0)*}_{\gamma} \widetilde{\mathcal M}^{(g)}_{\rm box}\bigg]}{\sum\limits_{spins}
\left|\mathcal{M}^{(0)}_\gamma\right|^2}
\nonumber\\
&=&\! \frac{2 \pi\alpha}{M}\left[\frac{Q^2}{Q^2+4EE^\prime}\right] 
{\mathcal R}e\,\left\{\frac{1}{i}\int \frac{{\rm d}^4 k}{{(2 \pi)}^4}\frac{{\rm Tr}[(\slashed{p}+m_l)\,\not{\!v}\,
(\slashed{p}^{\,\prime}+m_l)\,\slashed{v}\, (\slashed{p}-\slashed{k}+m_l)\, 
\slashed{v}]}{\left[(Q-k)^2+i0\right]\,(k^2-2 k\cdot p+i0)\,(v\cdot k+i0)^2}\right\}
\nonumber\\
\!&=&\! -\,\frac{4\pi\alpha}{M}\left[\frac{Q^2}{Q^2+4EE^\prime}\right] 
{\mathcal R}e\,\bigg\{(Q^2+8 E^2)\, I^{-}(p,0|0,1,1,1) - 2E(Q^2+4 E^2)\,  I^{-}(p,0|0,1,1,2) 
\nonumber \\
&& \hspace{4cm} -\, 2(E^\prime p + Ep^\prime) \cdot I^-_1(p,0|0,1,1,2)\bigg\}
\nonumber\\
\!&=&\! -\frac{4\pi\alpha}{M}\left[\frac{Q^2}{Q^2+4E^2}\right] 
{\mathcal R}e\, \bigg\{(Q^2+8 E^2)\, I^{(0)}(p,0|0,1,1,1) - E\left(Q^2+8E^2+4m^2_l\right)\, I^{(0)}(p,0|0,1,1,2) 
\nonumber \\
&& \hspace{3.7cm} -\, 2\left[(E^\prime p + Ep^\prime) \cdot T^{-}_1(p,0|0,1,1,2)\right]_{\rm LO} \bigg\} 
+ {\mathcal O}\left(\frac{1}{M^2}\right) \,,
\label{delta-g}
\end{eqnarray}
and 
\begin{eqnarray}
\delta^{(h)}_{\rm xbox}(Q^2) \!&=&\! \frac{2{\mathcal R}e\sum\limits_{spins}
\bigg[\mathcal{M}^{(0)}_{\gamma} \widetilde{\mathcal M}^{(h)}_{\rm xbox}\bigg]}{\sum\limits_{spins}
\left|\mathcal{M}^{(0)}_\gamma\right|^2}
\nonumber\\
&=&\! \frac{2 \pi\alpha}{M}\left[\frac{Q^2}{Q^2+4EE^\prime}\right]  
{\mathcal R}e\left\{\frac{1}{i}\int \frac{{\rm d}^4 k}{{(2 \pi)}^4}\frac{{\rm Tr}[(\slashed{p}+m_l)\,\slashed{v}\,
(\slashed{p}^{\, \prime}+m_l)\,\slashed{v}\, (\slashed{p}+\slashed{k}-\slashed{Q}+m_l)\, 
\not{\!v}]}{\left[(Q-k)^2+i0\right]\,(k^2+2 k\cdot p^{\prime}+i0)\,(v\cdot k+i0)^2}\right\}
\nonumber\\
\!&=&\! -\,\frac{4\pi\alpha}{M}\left[\frac{Q^2}{Q^2+4EE^\prime}\right] 
{\mathcal R}e\,\bigg\{\!-(Q^2+8 E^2)\,  I^{+}(p^{\prime},0|0,1,1,1) - 2E(Q^2+4 E^2)\,  I^{+}(p^{\prime},0|0,1,1,2) 
\nonumber \\
&& \hspace{4cm} +\, 2(E^\prime p + E p^\prime) \cdot I^+_1(p^\prime,0|0,1,1,2) \bigg\}
\nonumber \\
\!&=&\! -\frac{4\pi\alpha}{M}\left[\frac{Q^2}{Q^2+4E^2}\right] 
{\mathcal R}e\,\bigg\{(Q^2+8 E^2)\, I^{(0)}(p,0|0,1,1,1) + E\left(Q^2+8E^2+4m^2_l\right)\, I^{(0)}(p,0|0,1,1,2) 
\nonumber \\
&& \hspace{4cm} +\, 2\left[(E^\prime p + Ep^\prime) \cdot T^{+}_1(p^\prime,0|0,1,1,2)\right]_{\rm LO} \bigg\} 
\nonumber \\
&&\! -\, \frac{\alpha Q^2}{2\pi M E\beta^3}\left[\frac{Q^2+8E^2+4m^2_l}{Q^2+4E^2}\right] 
\left\{\ln\sqrt{\frac{1+\beta}{1-\beta}}-\beta\right\} 
+ {\mathcal O}\left(\frac{1}{M^2}\right)\, , 
\label{delta-h}
\end{eqnarray}
\end{widetext}
respectively, The above functions, $I^{-\mu}_1(p,0|0,1,1,2)$ and $I^{+\mu}_1(p^\prime,0|0,1,1,2)$ [or alternatively, 
$T^{-\mu}_1(p,0|0,1,1,2)$ and $T^{+\mu}_1(p^\prime,0|0,1,1,2)$], are rank-1 tensor integrals containing a single power
of the loop momentum $k^\mu$ [or alternatively, $(k-p)^\mu$ and $(k+p^\prime)^\mu$] in the numerators, 
Eqs.~\eqref{eq:I1_tensors} and ~\eqref{eq:T1_tensors}. The symbol $[...]_{\rm LO}$ in the above expressions denotes LO
terms to be considered within braces. We should note that when we add the (g) and (h) contributions from 
Eqs.~(\ref{delta-g}) and (\ref{delta-h}), the integral $I^{(0)}(p,0|0,1,1,2)$ cancels. Furthermore, only the difference,
$T_1^{-} - T_1^{+}$, is relevant when considering the sum of (g) and (h) diagrams.
The above integrals are evaluated by decomposing into simple scalar master integrals {\it via} the standard technique 
of {\it Passarino-Veltman reduction}~\cite{Passarino:1979} (PV). To this end, we can decompose the tensor structures of
$T^{\pm\, \mu}_1$ in terms of three independent external four-vectors, e.g., $v$, $p$ and $p^\prime$:
\begin{eqnarray}
T^{-\mu}_1(p,0|0,1,1,2) \!&=&\! v^\mu C^{-}_1 + p^{\prime\, \mu} C^{-}_2\,,
\nonumber\\
T^{+\mu}_1(p^\prime,0|0,1,1,2) \!&=&\! v^\mu C^{+}_1 + p^{\mu} C^{+}_2\,. 
\end{eqnarray}
The coefficients, $C^\pm_{1,2} = \,\stackrel{\circ}{C^\pm_{1,2}}+{\mathcal O}(1/M)$, are combinations of scalar master 
integrals, as discussed in Appendix~\ref{appB}. Only the LO parts of $C^\pm_{1,2}$ (as obtained by replacing $E^\prime=E$ 
and $\beta^\prime=\beta$) are relevant in our context, namely,
\begin{widetext}
\begin{eqnarray}
\stackrel{\circ}{C^{-}_1} \!&=&\! \left[1-\frac{1}{\beta^2}\right]I^{(0)}(p,0|0,1,1,1)  
+ \frac{1}{2E\beta^2}\left[I^{(0)}(p,0|0,0,1,2) - I(Q|0,1,0,2)\right]\,,
\nonumber\\
\nonumber\\
\stackrel{\circ}{C^{-}_2} \!&=&\! \frac{1}{E\beta^2}I^{(0)}(p,0|0,1,1,1) - I^{(0)}(p,0|0,1,1,2) 
- \frac{1}{2E^2\beta^2}\left[I^{(0)}(p,0|0,0,1,2) - I(Q|0,1,0,2)\right]\,,
\nonumber\\
\nonumber\\
\stackrel{\circ}{C^{+}_1} \!&=&\! -\,\left[1-\frac{1}{\beta^2}\right]I^{(0)}(p,0|0,1,1,1) 
- \frac{1}{2E\beta^2}\left[I^{(0)}(p,0|0,0,1,2) - I(Q|0,1,0,2)\right] \,, \quad \text{and}
\nonumber\\
\nonumber\\
\stackrel{\circ}{C^{+}_2} \!&=&\! -\,\frac{1}{E\beta^2}I^{(0)}(p,0|0,1,1,1) - I^{(0)}(p,0|0,1,1,2) 
+ \frac{1}{2E^2\beta^2}\left[I^{(0)}(p,0|0,0,1,2) - I(Q|0,1,0,2)\right]
\nonumber\\
&&\! -\, \frac{2}{(4\pi)^2 E^2 \beta^3} \left[\ln \sqrt{\frac{1+\beta}{1-\beta}}-\beta\right] \,. 
\label{eq:C12}
\end{eqnarray}  
\end{widetext}
In particular, the two-point functions, $I^{(0)}(p,0|0,0,1,2)$ and $I(Q|0,1,0,2)$, appear as a difference in our chiral
order. Therefore, their UV divergences cancel exactly as they should since we only take the difference of these integrals
in each of the above coefficients $\stackrel{\circ}{C^\pm_{1,2}}$, see Eq.~(\ref{eq:C12}) or Eq.~(\ref{eq:C12inB}):
\begin{eqnarray}
&& I^{(0)}(p,0|0,0,1,2) - I(Q|0,1,0,2) 
\nonumber\\
&&\hspace{1cm} =\, \frac{1}{8 \pi^2} \left[  \frac{2}{\beta} \ln \sqrt{\frac{1+\beta}{1-\beta}} 
+ \ln\left(\frac{ M^2 m_l^2}{Q^4}\right) \,\,\right]\,.\qquad\, 
\label{I0-IQ}
\end{eqnarray}
All our results for the TPE radiative corrections are UV-finite, as expected from naive dimensional arguments. We observe 
that the presence of the lepton mass-dependent logarithmic term $\propto \ln\left(\frac{ M^2 m_l^2}{Q^4}\right)$ of 
Eq.~\eqref{I0-IQ}, originating 
from the (g) and (h) diagrams, enhance the contributions for the electron scattering as 
compared to muon scattering. 
(see Fig.~\ref{LO-NLO-results-1}).  

Furthermore, it is important to note that the three-point function $I^{(0)}(p,0|0,1,1,2)$ appearing in the 
individual contributions from the (g) and (h) diagrams, $\delta^{(g)}_{\rm box}$ and $\delta^{(h)}_{\rm xbox}$, is  
linear in the proton's mass $M$ [cf. Eqs.~\eqref{eq:I-0112}], and therefore, may lead to pathologies with the 
convergence of the chiral expansion. However, as noted, this integral appearing in both $\delta^{(g)}_{\rm box}$ 
and $\delta^{(g)}_{\rm xbox}$ cancel in the sum. Our combined result for the fractional contribution from the (g) 
and (h) diagrams become
\begin{widetext}
\begin{eqnarray}
\delta^{(gh)}_{\gamma\gamma}(Q^2) \!&=&\!  \delta^{(g)}_{\rm box}(Q^2) + \delta^{(h)}_{\rm xbox}(Q^2)
\nonumber\\
&=&\! -\,\frac{4\pi\alpha}{M}\left[\frac{Q^2}{Q^2+4E^2}\right] {\mathcal R}e \,
\Bigg\{ 2\left[Q^2\left(1+\frac{1}{\beta^2}\right)+8E^2\right] I^{(0)}(p,0|0,1,1,1) 
\nonumber \\
&&\hspace{3.8cm} -\, \left(\frac{Q^2+4E^2\beta^2}{E\beta^2}\right)\left[I^{(0)}(p,0|0,0,1,2) - I(Q|0,1,0,2)\right] \Bigg\}
\nonumber\\
&&\! -\, \frac{\alpha Q^2}{\pi M E\beta^3}\left[\ln\sqrt{\frac{1+\beta}{1-\beta}}-\beta\right] 
+ {\mathcal O}\left(\frac{1}{M^2}\right)\,.
\end{eqnarray}

\begin{figure*}
\begin{center}
		\includegraphics[scale=0.58]{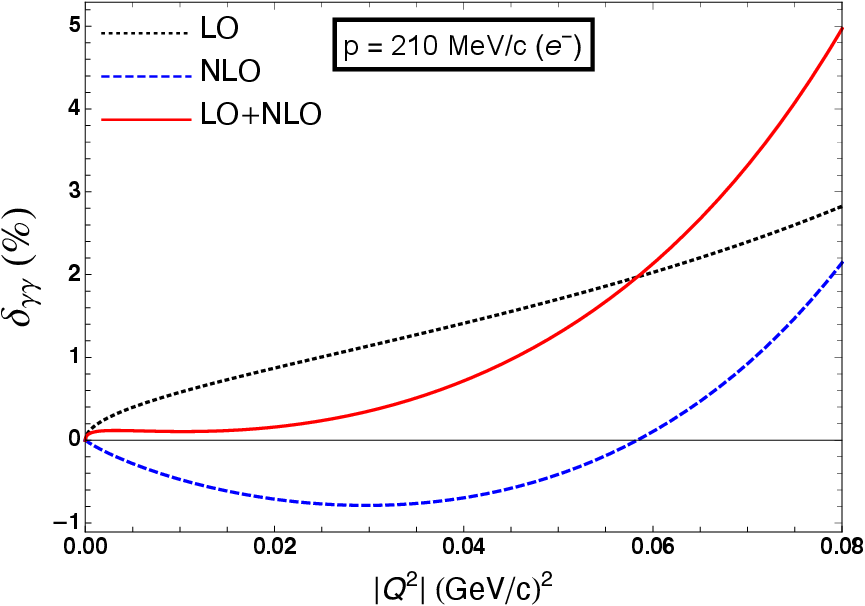}\qquad
		\includegraphics[scale=0.58]{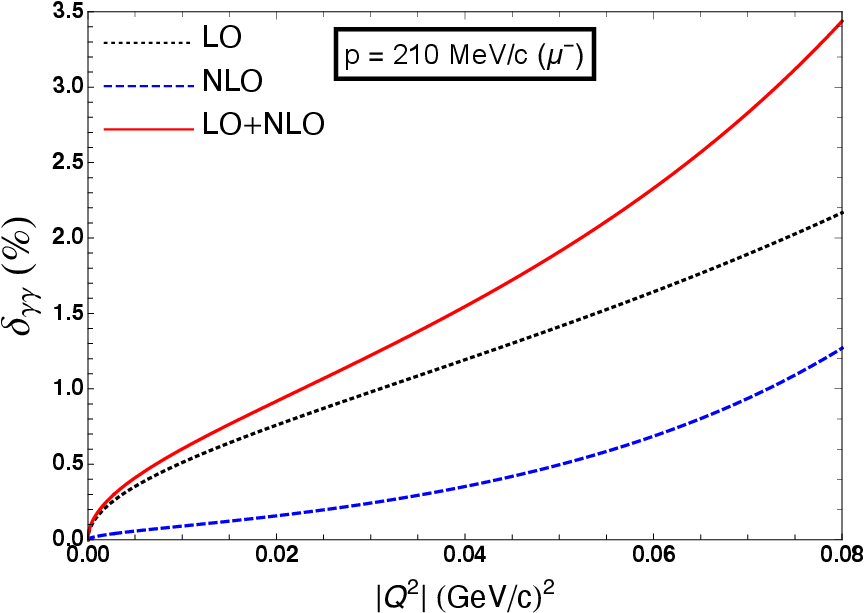} \\
  
        \vspace{1.5cm}
        
        \includegraphics[scale=0.58]{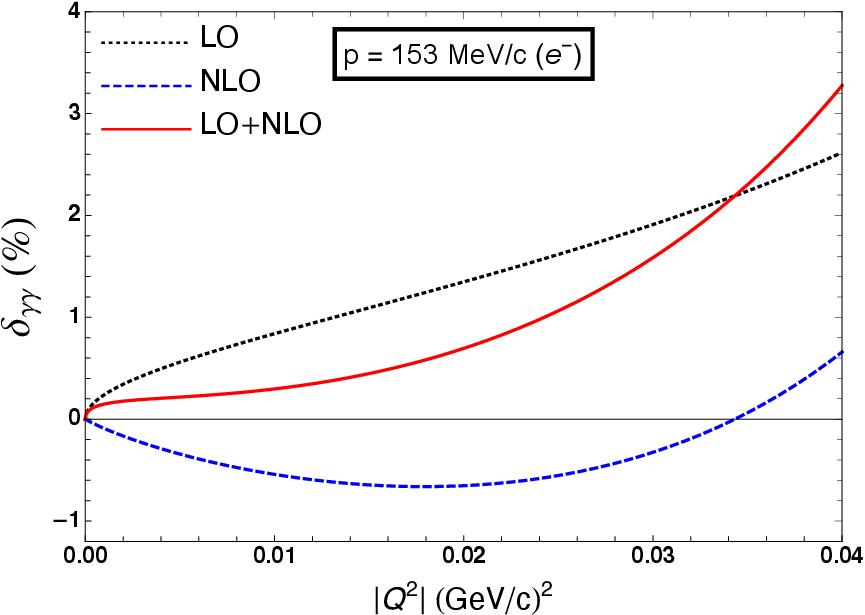}\qquad
		\includegraphics[scale=0.58]{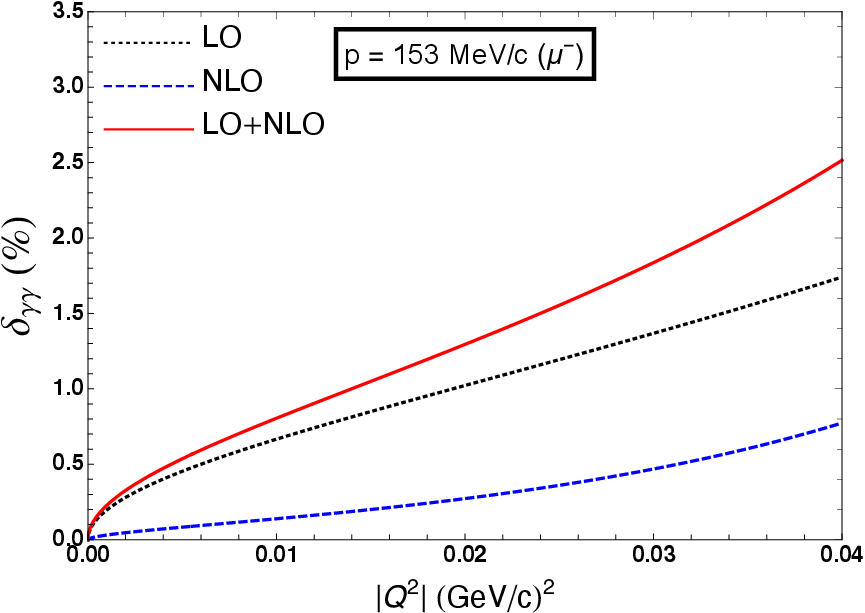} \\
  
       \vspace{1.5cm}
       
        \includegraphics[scale=0.58]{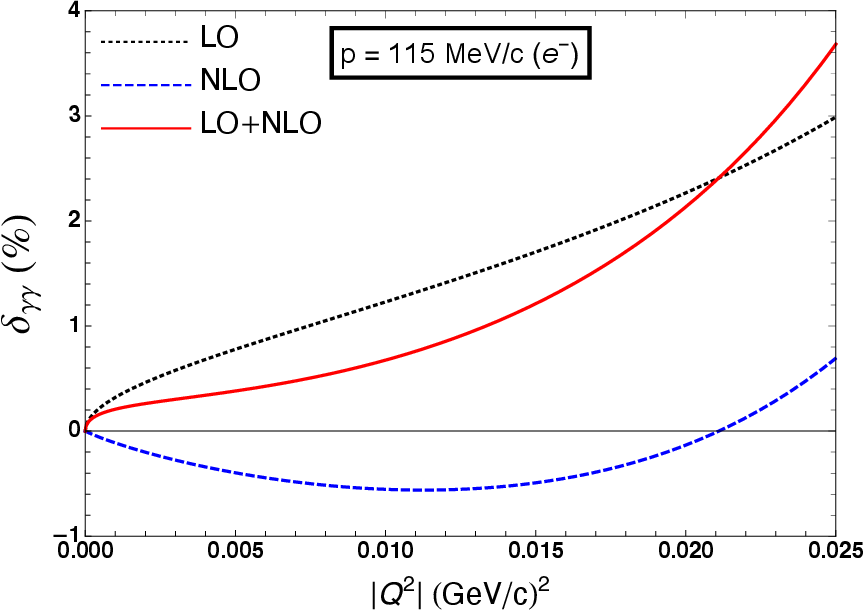}\qquad
		\includegraphics[scale=0.58]{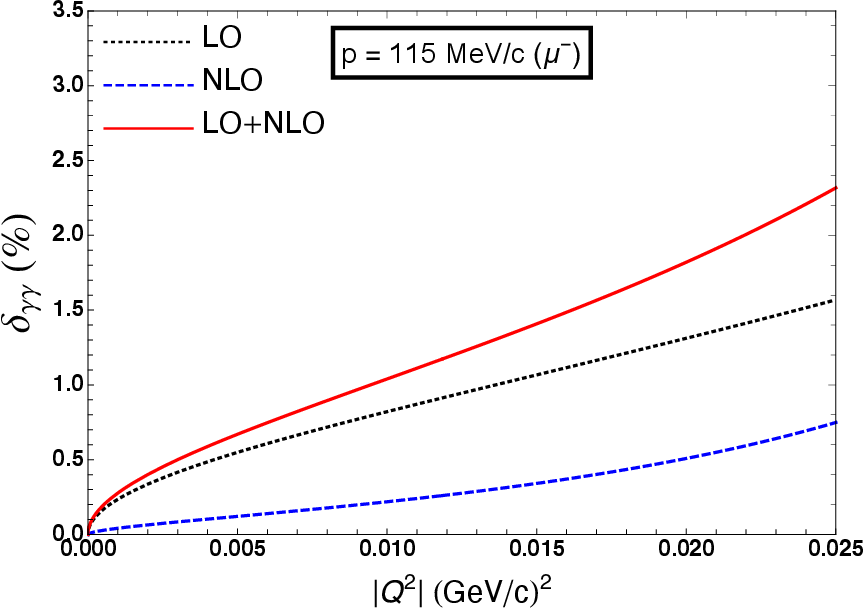} 
		\caption{The fractional TPE contributions (expressed in percentage) to the elastic lepton-proton scattering 
                 cross-section [i.e., of ${\mathcal O}(\alpha)$] in HB$\chi$PT. Left (right) plots show TPE 
                 contribution for e-p ($\mu$-p) scattering,  respectively, at specific MUSE beam momenta. Each plot 
                 covers the full kinematical scattering range, $0<|Q^2|<|Q^2_{\rm max}|$  with $\theta\in[0,\pi]$. 
                 The dotted (black) lines correspond to LO results, the dashed (blue) lines correspond to NLO 
                 corrections, while the thick solid (red) lines denote the total (LO plus NLO) contributions from 
                 all the TPE diagrams displayed in Fig.~\ref{fig1}.}
\label{LO-NLO-results-2}
\end{center}
\end{figure*}

Finally, the least important NLO contribution to the TPE radiative corrections arises from the seagull diagram (i), 
without an elastic proton intermediate state. After initial cancellations, only the residual part 
$\widetilde{\mathcal M}^{(i)}_{\rm seagull}$ [see Eq.~\eqref{17}] proportional to $g^{\mu\nu}$ contributes to the 
NLO cross-section: 
\begin{eqnarray}
\delta^{\rm (seagull)}_{\gamma\gamma} (Q^2) \!&=&\! 
\frac{2{\mathcal R}e\,\sum_{spins}
\bigg[{\mathcal M}^{(0)*}_\gamma\,\widetilde{\mathcal M}^{(i)}_{\rm seagull}\bigg]}{\sum_{spin}
\left|{\mathcal M}^{(0)}_\gamma\right|^2}
\nonumber\\
&=&\!  \frac{4\pi\alpha}{M}\left[\frac{Q^2}{Q^2+4EE^\prime}\right]  
{\mathcal R}e\,\left\{\frac{1}{i}\int \frac{{\rm d}^4 k}{{(2 \pi)}^4}\frac{{\rm Tr}[(\not{\!p}+m_l)\,\gamma^\rho\,
(\not{\!p^{\prime}}+m_l)\,\gamma^\mu \,(\not{\!p}-\not{\!k}+m_l)\,
\gamma_\mu \,v_\rho]}{(k^2+i0)\,\left[(Q-k)^2+i0\right]\,(k^2-2 k\cdot p+i0)}\right\} 
\nonumber \\
&=&\, \frac{4\pi\alpha}{M}\left\{\frac{Q^2}{Q^2+4EE^\prime}\right] 
{\mathcal R}e\,\bigg\{4\left(Q^2 v+2 E p+2 E p^{\prime}\right)\cdot I^{-}_1(p,0|1,1,1,0) + 16 m_l^2 E\, I(Q|1,1,1,0)\bigg\} 
\nonumber 
\end{eqnarray}
\begin{eqnarray}
&=&\! -\frac{4 \alpha E}{\pi M}\left[\frac{Q^2}{Q^2+4E^2}\right]
{\mathcal R}e\, \bigg\{\left(\frac{1-\nu^2}{2 \nu^2}\right) \ln\left(\frac{-Q^2}{m^2_l}\right) 
- (4\pi)^2m_l^2\left(\frac{1+\nu^2}{\nu^2}\right) I(Q|1,1,1,0) \bigg\} + {\mathcal O}\left(\frac{1}{M^2}\right)\,, 
\label{eq:seagull}
\end{eqnarray}
\end{widetext}
where $\nu=\sqrt{1-4m^2_l/Q^2}$ is a $Q^2$-dependent kinematical variable. The analytical results for the relativistic
three-point scalar $I(Q|1,1,1,0)$ and tensor $I^{-\mu}_1(p,0|1,1,1,0)$ integrals are expressed in Eqs.~\eqref{I1110} 
and \eqref{seagull-tensor}. In the case of muon scattering, the second term in the seagull contribution being 
proportional to the square of its mass leads to some enhancements in the result as compared to the electron case, where
the contribution is tiny (see Fig.~\ref{LO-NLO-results-1}). For a detailed discussion about the seagull contribution, 
we refer the reader to Ref.~\cite{Talukdar:2019dko}.

\section{Results and discussion}\label{Results}
In Figs.~\ref{LO-results} - \ref{LO-NLO-results-2}, we provide the numerical results of our analytically derived 
expressions for the box, crossed-box, and seagull TPE contributions, after removal of the IR-divergent terms [see 
Eq.~\eqref{deltaIRbox}].\footnote{The exact cancellation of the IR-divergent terms from the TPE diagrams that we 
have derived, namely, $\delta^{\rm (box)}_{\rm IR}$, with the corresponding soft-bremsstrahlung counterpart, 
$\delta^{\rm (soft)}_{\rm IR}$, is a concrete result that we have established {\it via} explicit HB$\chi$PT 
calculations including NLO corrections to the latter contributions. The details of such a calculation will be 
presented in a follow-up publication concerning the TPE contribution to charge asymmetry.} In Fig.~\ref{LO-results}
our LO correction results, Eq.~(\ref{eq:ab-LO}), are plotted for e-p and $\mu$-p scatterings, at the three specific
MUSE choices of the incoming lepton momenta. In the same figure, we also compared our results to other TPE works, 
e.g., Ref.~\cite{Koshchii:2017dzr}, based on relativistic QED invoking SPA~\cite{Tsai:1961zz}, as well as the 
well-known estimate, Eq.~\eqref{Feshbach} of McKinley and Feshbach~\cite{McKinley:1948zz}, based on the scattering
of relativistic electrons off static Coulomb potentials in the context of the second Born approximation.  

Our final results of evaluating the TPE radiative corrections to the elastic cross-sections are consolidated in 
Fig.~\ref{LO-NLO-results-2}, where we present a comparison of LO versus NLO  contribution from HB$\chi$PT. As seen 
from the figure [see also Fig.~\ref{LO-LO1/M-results}], for e-p scattering the HB$\chi$PT expansion does not appear 
to be reliable. In the figures, LO corresponds to the {\it true} LO result which stems from diagrams (a) and (b), 
Eq.~\eqref{eq:ab-LO}. In addition, as we have discussed earlier, the diagrams (a) and (b) also contain NLO 
contributions from Eq.~\eqref{delta-ab2} that for the case of e-p scattering are dominating the positive LO part. 
In fact, these NLO contributions from the (a) and (b) diagrams make the total NLO corrections from all TPE diagrams, 
(a) - (i) in Fig.~\ref{fig1} negative for the most part of the $Q^2$ domain, as shown in the left panel plots of
Fig.~\ref{LO-NLO-results-2}. However, in a forthcoming publication, we shall demonstrate that our NLO results diminish  
when we consider the contributions of the finite part of the corresponding soft-bremsstrahlung diagrams. We shall show
that the source of this change in our results can be attributed to the ``large" ${\rm ln}\left(-Q^2/m_l^2 \right)$ 
which appears in the soft-bremsstrahlung process. In contrast, for muon-proton scattering with $Q^2\sim m^2_\mu$, the 
above  "log"-factor is very small and will not markedly affect the HB$\chi$PT perturbative expansion. Another 
observation concerns the NLO propagator insertion diagrams, (g) and (h), whose contributions are almost as large as 
the LO contribution for e-p scattering. The former corrections have a positive sign and they diminish the negative NLO
corrections from (a) and (b) diagrams, which also include $1/M$ order correction terms to the proton propagator. We 
further observe that the box diagram (c) exactly cancels the contribution from the crossed-box (d) diagram, see below
Eq.~\eqref{delta_c+d}. Finally, we find from Fig.~\ref{LO-NLO-results-2} that for e-p scattering the two IR-finite 
diagrams, (e) and (f), become large for $Q^2 \gtrsim 0.02$ (GeV/c)$^2$, especially due to the contribution of the 
relativistic three-point integral $I(Q|1,1,1,0)$. In this context, we note that even at the kinematic regime relevant 
to the MUSE, the electron is relativistic. However, the muon being much heavier is not expected to behave as a true 
relativistic probe in this regime.

Turning to muon-proton scattering results we find significant differences in the TPE corrections compared to the 
electron-proton scattering results. For the $\mu$-p scattering case, the LO contribution dominates the NLO 
contribution in the entire MUSE kinematical domain. Furthermore, both LO and NLO contributions are positive in the 
whole $Q^2$ range. These results indicate that the perturbative aspects of HB$\chi$PT appear to behave more robustly 
for $\mu$-p scattering. In particular, we find from Fig.~\ref{LO-NLO-results-1} (right panel) that the contributions 
from the NLO TPE diagrams (g) and (h) are significantly smaller than that of the LO, which sharply contrasts the 
corresponding e-p scattering results (left panel). Nonetheless, their contributions are the most significant ones 
among all other NLO corrections. Another contrasting feature regarding the muon results is the following: while on the
one hand, the (e) and (f) diagrams have rather small contributions in this case, the role of the seagull diagram, on 
the other hand, becomes quite relevant, see Fig.~\ref{LO-NLO-results-1}. Finally, we observe that the total TPE 
correction including the NLO contributions to $\mu$-p scattering is somewhat smaller than the size of e-p scattering.

We can clearly see the impact of our exact analytical results in comparison to the SPA results of 
Ref.~\cite{Talukdar:2019dko} also in the context of HB$\chi$PT. There the calculation applied SPA both in the numerator
and denominator of the lepton propagators which appear in the TPE box and crossed-box diagrams. The results in that work 
were expressed solely in terms of the IR-divergent three-point functions, $I^{-}(p,0|1,0,1,1)$ and 
$I^{+}(p^\prime,0 |1,0,1,1)$. The application of SPA is in effect tantamount to suppressing the effects arising from 
several such integrals which potentially yield sizeable numerical contributions, as evidenced in 
Figs.~\ref{LO-NLO-results-1} and \ref{LO-NLO-results-2}. For instance, the (e), (f), (g), and (h) diagram contributions 
would be greatly diminished in SPA by the absence of integrals, such as the three-point function, $I(Q|1,1,1,0)$, and to 
some extent the four-point function, $Z^\pm$. The most significant difference in the SPA results compared with our exact 
result is, however, the vanishing true LO contribution from  diagrams (a) and (b) as obtained in 
Ref.~\cite{Talukdar:2019dko}. In other words, Ref.~\cite{Talukdar:2019dko} completely missed out on the Feshbach 
contribution, Eq.~\eqref{Feshbach}~\cite{McKinley:1948zz}. In contrast, our current results in Fig.~\ref{LO-results} 
suggest that the LO contribution to TPE radiative corrections exceeds $2\%$ at the largest $Q^2$ values, essentially 
stemming from the integral $I(Q|1,1,0,1)$ which is absent in a SPA evaluation. Nevertheless, a notable feature of the 
TPE radiative corrections is that at $Q^2=0$ these contributions identically vanish in all the different 
approaches/approximations.

\section{Summary and conclusion}\label{Summary}
We have presented in this work an exact analytical evaluation of LO and NLO contributions from two-photon exchange to the
lepton-proton elastic unpolarized cross-section at low-energies in HB$\chi$PT, taking nonzero lepton mass into account.
We find sizeable contributions beyond the expected SPA results, akin to the ones found {\it via} dispersion techniques, e.g., 
Ref.~\cite{Tomalak:2014sva}. Our EFT method contrasts several prior TPE analyses utilizing relativistic hadronic models, 
which frequently parameterize the proton-photon vertices using phenomenological form factors. It is notable that as per the 
tenets of HB$\chi$PT power counting, the proton is a point particle at order $1/M$. The proton structure effects, e.g., 
anomalous magnetic moment and pion-loop contributions, only start appearing at order $1/M^2$. Although such NNLO effects will
also appear in the usual Born differential cross-section 
$\left[{\rm d} \sigma_{el}(Q^2)/{\rm d} \Omega^\prime_l \right]_{\gamma}$, their influence on the fractional TPE radiative 
corrections $\delta^{\rm (box)}_{\gamma\gamma}$ are certainly sub-dominant. In fact, Ref.~\cite{Kaiser:2010zz} has done an 
exact e-$\mu$ TPE scattering calculation in the context of standard relativistic QED. Using the e-$\mu$ result of 
Ref.~\cite{Kaiser:2010zz}, in principle, one can replace the muon mass with the proton mass $M$ and make an expansion in 
$1/M$ and find our HB$\chi$PT $1/M$ results. However, such a calculation requires a nontrivial $1/M$ expansion which we have not
performed in this work. The paper solely deals with the exact NLO estimation of the TPE radiative corrections without 
proton's form factors modification in a non-systematic manner (although it is well-known that form factors reduce the overall 
magnitude of the TPE corrections). Nevertheless, a systematic NNLO calculation of the TPE should be pursued as our future 
endeavor. The major difference with the SPA evaluation of Ref.~\cite{Talukdar:2019dko} is that our calculations are done 
exactly including all soft- and hard-photon exchange configurations of the TPE loops. In fact, our current results are expected
to exhibit sizeable differences with any existing SPA calculation using a perturbative/diagrammatic approach. Regarding the 
isolation of the IR divergences, we used dimensional regularization. They arise from the three-point functions, 
$I^{-}(p,0|1,0,1,1)$ and $I^{+}(p^\prime,0 |1,0,1,1)$ stemming only from diagrams (a) and (b). The remaining TPE diagrams have 
all finite contributions. Such an IR singular structure of our TPE results agrees with the recent evaluation base of manifestly 
Lorentz-invariant B$\chi$PT~\cite{Cao:2021nhm}. In our HB$\chi$PT approach, the IR singularities appearing at LO automatically 
drop out of the calculation with only residual NLO IR-singular terms remaining and they cancel against the relevant 
soft-bremsstrahlung IR-singular contributions at NLO.

\section*{ACKNOWLEDGMENTS}
DC and PC acknowledge the partial financial support from the Science and Engineering Research Board under 
grant number CRG/2019/000895. UR acknowledges partial financial support from the Science and Engineering 
Research Board under grant number CRG/2022/000027, and US Department of Energy, Award No. DE-FG02-93ER-40756.
UR is also grateful to the Department of Physics and Astronomy, University of South Carolina, Columbia, and
the Institute of Nuclear and Particle Physics, Ohio University, Athens, for their local hospitality and 
support during the completion stages of this work. The authors are grateful to Steffen Stauch and Dipangkar 
Dutta for useful discussions. Last but not least, special thanks go to Rakshanda Goswami, Bheemsehan Gurjar, 
Bhoomika Das, and Ghanashyam Meher, for helping out with cross-checking many of our calculations.

\appendix
\section{Notations}\label{appA}
In this section, we present the notations for all the Feynman loop integrals used in this work. First, we 
start by presenting the following generic forms of the one-loop scalar master integrals with external 
four-momenta $p$ and $p^\prime$, four-momentum transfer $Q_\mu=(p-p^\prime)_\mu$, and loop 
four-momentum $k_\mu$, having indices $n_{1,2,3,4}\in {\mathbb Z}_\pm$ and a real-valued parameter 
$\omega$:
\begin{widetext}
\begin{eqnarray}
I^-(p,\omega|n_1,n_2,n_3,n_4) \!&=&\! 
\frac{1}{i}\int \frac{{\rm d}^4 k}{{(2 \pi)}^4} 
\frac{1}{(k^2+i 0)^{n1}\,[(k-Q)^2+i0]^{n2}\,(k^2 - 2k\cdot p+i0)^{n_3}\, (v\cdot k+\omega+i0)^{n4}}\,,
\nonumber\\
\nonumber\\
I^+ (p^\prime,\omega|n_1,n_2,n_3,n_4) \!&=&\! 
\frac{1}{i}\int \frac{{\rm d}^4 k}{{(2 \pi)}^4} 
\frac{1}{(k^2+i 0)^{n1}\,[(k-Q)^2+i0]^{n2}\,(k^2 + 2k\cdot p^\prime+i0)^{n_3}\, (v\cdot k+\omega+i0)^{n4}}\,,
\nonumber\\
\nonumber\\
Z^-(\Delta,i\sqrt{-Q^2}/2,m_l,\omega) \!&=&\! 
\frac{1}{i}\int \frac{{\rm d}^4 k}{(2 \pi)^4} 
\frac{1}{\left[(k+\Delta)^2-\frac{1}{4}Q^2+i0\right]\,(k^2-m^{2}_l+i0)\,\left(v\cdot k+\omega+i0\right)}\,, 
\quad \text{and}
\nonumber\\ 
\nonumber\\
Z^+(\Delta^\prime,i\sqrt{-Q^2}/2,m_l,\omega) \!&=&\! 
\frac{1}{i}\int \frac{{\rm d}^4 k}{(2 \pi)^4} 
\frac{1}{\left[(k+\Delta^\prime)^2-\frac{1}{4}Q^2+i0\right]\,(k^2-m^{2}_l+i0)\,\left(v\cdot k+\omega+i0 \right)}\,,
\label{eq:generic_scalar}
\end{eqnarray} 
\end{widetext}
where the $Z^\pm$ integrals are functions of the additional four vectors, 
$\Delta_\mu=\left(p-\frac{Q}{2}\right)_\mu$ and 
$\Delta^\prime_\mu=-\left(p^\prime+\frac{Q}{2}\right)_\mu=-\left(p-\frac{Q}{2}\right)_\mu=-\Delta_\mu$. 
The parameter $\omega$ can be either zero,  $\omega=v\cdot Q=-\frac{Q^2}{2M}$, or any finite quantity (e.g., 
$\omega=v\cdot p=E\,,\,\, -v\cdot p^\prime=-E^\prime$), depending on the specific integrals we are dealing 
with.

Next, we have the two tensor integrals with one power of the loop momentum in the numerators appearing in 
our work. They have the following generic forms:
\begin{widetext}
\begin{eqnarray}
I^{-\mu}_1(p,\omega|n_1,n_2,n_3,n_4)\! 
&=&\! \frac{1}{i}\int \frac{{\rm d}^4 k}{{(2 \pi)}^4} 
\frac{k^\mu}{(k^2+i 0)^{n1}\,[(k-Q)^2+i0]^{n2}\,(k^2- 2k\cdot p+i0)^{n_3}\, (v\cdot k+\omega+i0)^{n4}}\,, \quad \text{and}
\nonumber\\
\nonumber\\
I^{+\mu}_1(p^\prime,\omega|n_1,n_2,n_3,n_4)\!
&=&\! \frac{1}{i}\int \frac{{\rm d}^4 k}{{(2 \pi)}^4} 
\frac{k^\mu}{(k^2+i 0)^{n1}\,[(k-Q)^2+i0]^{n2}\,(k^2+ 2k\cdot p^\prime+i0)^{n_3}\, (v\cdot k+\omega+i0)^{n4}}\,.
\label{eq:I1_tensors}
\end{eqnarray}
In particular, using the following transformations on the tensor integrals, $I^{-\mu}_1(p,0|0,1,1,2)$ and 
$I^{+\mu}_1(p^\prime,0|0,1,1,2)$, used in this work:
\begin{eqnarray}
2(E^\prime p+E p^\prime)\cdot I^{-}_1(p,0|0,1,1,2) \!&=&\!  2(E^\prime p + E p^\prime)\cdot T^{-}_1(p,0|0,1,1,2) 
- \left[EQ^2-2m^2_l(E+E^\prime)\right] I^{-}(p,0|0,1,1,2)\,,\,\, \text{and}
\nonumber\\
\nonumber\\
2(E^\prime p + Ep^\prime)\cdot I^{+}_1(p^\prime,0|0,1,1,2) \!&=&\!  2(E^\prime p + Ep^\prime)\cdot T^{+}_1(p,0|0,1,1,2) 
+ \left[EQ^2-2m^2_l(E+E^\prime)\right]I^{+}(p^\prime,0| 0,1,1,2)\,,
\label{eq:I1-T1_tranform}
\end{eqnarray}
we obtain two more tensor integrals of interest:
\begin{eqnarray}
T^{-\mu}_1(p,0|0,1,1,2) \!&=&\! \frac{1}{i}\int \frac{{\rm d}^4 k}{{(2 \pi)}^4} 
\frac{(k-p)^\mu}{[(k-Q)^2+i0]\,(k^2- 2k\cdot p+i0)\,(v\cdot k +i0)^2}\,,
\nonumber\\
\nonumber\\
T^{+\mu}_1(p^\prime,0|0,1,1,2) \!&=&\! \frac{1}{i}\int \frac{{\rm d}^4 k}{{(2 \pi)}^4} 
\frac{(k+p^\prime)^\mu}{[(k-Q)^2+i0]\,(k^2+ 2k\cdot p^\prime+i0)\, (v\cdot k+i0)^{2}}\,. 
\label{eq:T1_tensors} 
\end{eqnarray}
\end{widetext}

\section{Analytical results for master integrals}\label{appB}
Before we display our analytical expressions for the scalar master integrals, we briefly discuss the reduction of the 
rather complicated four-point functions, $I^-(p,0|1,1,1,1)$ and $I^-(p^\prime,0|1,1,1,1)$, appearing in TPE box and 
crossed-box diagrams, respectively, into simpler three- and four-point functions {\it via} the method of partial 
fractions. For this purpose, we first express the four-point integrals $I^\pm$ as
\begin{widetext}
\begin{eqnarray}
I^-(p,0|1,1,1,1) = \frac{1}{i}\int \frac{{\rm d}^4 k}{(2\pi)^4} \frac{1}{D_1 D_2 D_3 D_4}\,,\qquad \text{and} \qquad
I^+(p^\prime,0|1,1,1,1) = \frac{1}{i} \int \frac{{\rm d}^4 k}{(2\pi)^4} \frac{1}{D_1 D_2 \widetilde{D_3} D_4}\,,
\end{eqnarray}
where for brevity we have defined the following propagator denominators:
\begin{eqnarray}
D_1 = k^2+i0\,,\quad D_2 = (k-Q)^2+i0\,,\quad   D_3 = k^2-2 k \cdotp p+i0\,,\quad 
\widetilde{D_3} = k^2 + 2 k \cdotp p^\prime +i0\,,\quad \text{and}\quad D_4 = v \cdot k +i0\,.\,\quad
\end{eqnarray}  
Next, we use the method of partial fractions to decompose each of the above scalar four-point functions into 
two scalar three-point functions and one tensor four-point function:
\begin{eqnarray}
\frac{1}{i} \int \frac{{\rm d}^4 k}{(2\pi)^4} \frac{1}{D_1 D_2 D_3 D_4} \!&=&\! 
\frac{1}{i} \int \frac{{\rm d}^4 k}{(2\pi)^4}\, \frac{1}{Q^2} \left( \frac{1}{D_1 D_3 D_4}+\frac{1}{D_2 D_3 D_4} 
- \frac{2 k\cdot(k-Q)} {D_1 D_2 D_3 D_4}\right)\,, \quad \text{and}
\nonumber\\
\nonumber\\
\frac{1}{i} \int \frac{{\rm d}^4 k}{(2\pi)^4} \frac{1}{D_1 D_2 \widetilde{D_3} D_4} \!&=&\! 
\frac{1}{i} \int \frac{{\rm d}^4 k}{(2\pi)^4}\, \frac{1}{Q^2} \left( \frac{1}{ {D_1 \widetilde{D_3} D_4}} 
+\frac{1}{ {D_2 \widetilde{D_3} D_4}} - \frac{2 k\cdot (k-Q)} {D_1 D_2 \widetilde{D_3} D_4}\right)\,,
\label{I+1111_a}
\end{eqnarray}
which explicitly boil down to the results:
\begin{eqnarray}
I^-(p,0|1,1,1,1) \!&=&\! 
\frac{1}{i}\int \frac{{\rm d}^dk}{(2\pi)^d}\frac{1}{(k^2+i0)\,[(k-Q)^2+i0]\,(k^2-2k\cdot p+i0)\,(v\cdot k +i0)} 
\nonumber \\
&=&\! \,\frac{1}{Q^2}\Big[ I^-(p,0|1,0,1,1)+I^-(p,0|0,1,1,1)-2Z^-(\Delta,i\sqrt{-Q^2}/2,m_l,E)\Big]\,, \quad \text{and}
\nonumber\\
\nonumber\\
I^+(p^\prime,0|1,1,1,1) \!&=&\! 
\frac{1}{i}\int \frac{{\rm d}^4 k}{(2\pi)^4}\frac{1}{(k^2+i0)\,[(k-Q)^2+i0]\,(k^2+2k\cdot p^\prime +i0)\,(v\cdot k +i0)} 
\nonumber \\
&=&\! \frac{1}{Q^2}\Big[I^+(p^\prime,0|1,0,1,1)+I^+(p^\prime,0|0,1,1,1)-2Z^+(\Delta^\prime,i\sqrt{-Q^2}/2,m_l,-E^\prime)\Big]\,.
\label{I+1111_b}
\end{eqnarray}
Here we note that the tensor-like forms of the four-point functions $Z^\pm$ which follow from Eq.~\eqref{I+1111_a} 
can be conveniently transformed into the corresponding scalar forms {\it via} simple transformations of the loop 
momentum $k_\mu$:
\begin{eqnarray}
Z^{-}(p,0|1,1,1,1) \!&=&\! \frac{1}{i}\int \frac{{\rm d}^4 k}{(2\pi)^4}
\frac{ k\cdot (k-Q)}{(k^2+i0)\,[(k-Q)^2+i0]\,(k^2-2k\cdot p+i0)\,(v\cdot k +i0)}
\nonumber\\
&\stackrel{k\to k+p}{=}&\! \frac{1}{i}\int \frac{{\rm d}^4 k}{(2 \pi)^4}
\frac{1}{\left[(k+\Delta)^2-\frac{1}{4}Q^2+i0\right](k^2-m^{2}_l+i0)\left(v\cdot k+E+i0\right)}
\nonumber\\
&\equiv&\! Z^{-}(\Delta,i\sqrt{-Q^2}/2,m_l,E)\,, 
\end{eqnarray}
and similarly,
\begin{eqnarray}
Z^{+}(p^\prime,0|1,1,1,1) \!&=&\! \frac{1}{i}\int \frac{{\rm d}^4 k}{(2\pi)^4}
\frac{ k\cdot (k-Q)}{(k^2+i0)\,[(k-Q)^2+i0]\,(k^2+2k\cdot p^\prime+i0)\,(v\cdot k +i0)}
\nonumber\\
&\stackrel{k\to k-p^\prime}{=}&\! \frac{1}{i}\int \frac{{\rm d}^4 k}{(2 \pi)^4}
\frac{1}{\left[(k+\Delta^\prime)^2-\frac{1}{4}Q^2+i0\right](k^2-m^{2}_l+i0)\left(v\cdot k-E^\prime+i0\right)}
\nonumber\\
&\equiv&\! Z^{+}(\Delta^\prime,i\sqrt{-Q^2}/2,m_l,-E^\prime)\,.
\end{eqnarray}
\end{widetext}
The three- and four-point functions resulting from the above decompositions are individually rather 
straightforward to evaluate. While the three-point functions, $I^{-}(p,0|1,0,1,1)$ and 
$I^{+}(p^\prime,0|1,0,1,1)$ are IR-divergent, the three-point functions, 
$I^{-}(p,0|0,1,1,1)$ and $I^{+}(p^\prime,0|0,1,1,1)$, as well as the four-point functions, 
$Z^-(\Delta,i\sqrt{-Q^2}/2,m_l,E)$ and $Z^+(\Delta^\prime,i\sqrt{-Q^2}/2,m_l,-E^\prime)$, are all IR-finite. 
Each of these integrals contains one non-relativistic heavy baryon (proton) propagator linear in the loop 
four-momentum factor, namely, $v \cdot k$. 

We first focus on the IR-divergent master integrals, $I^{-}(p,0|1,0,1,1)$ and $I^{+}(p^\prime,0|1,0,1,1)$. 
One method of isolating the IR-divergent parts of such integrals is to utilize dimensional regularization 
by analytically continuing the integrals to $D$-dimensional ($D = 4 -2\epsilon$) space-time, where the pole
$\epsilon < 0$ yields the IR divergence. Especially, to deal with the non-relativistic propagator it is
convenient to employ a special form of the Feynman parameterization~\cite{zupan:2002}, namely,
\begin{eqnarray}
\left[\bigg(\prod\limits_{i=1}^N A_{i}\bigg)B\right]^{-1} \!\!&=&\!\! N \int_{0}^\infty 2  
{\rm d} \lambda \int \prod\limits_{i=1}^N {\rm d}u_i 
\nonumber\\
&&\! \times\,\frac{\delta (1-\sum_{i=1}^N u_i) \prod_{i=1}^n \theta(u_i)}{\bigg[\sum_{i=1}^N A_i u_i+2 B\lambda \bigg]^{N+1}} \,.
\nonumber\\
\label{zupan_para}
\end{eqnarray}
Using such a parameterization one can evaluate n-point Feynman integrals with one or more non-relativistic 
propagators. The detailed derivation of the Feynman integrals will not be presented here. Below, we simply 
display the analytical expressions of the specific integrals used in this work. We now explicitly spell out
the results for the IR-singular master integrals with an arbitrary choice of the renormalization scale $\mu$: 
\begin{widetext}
\begin{eqnarray}
	I^-(p,0|1,0,1,1) \!\!&\equiv&\!\!  I^{(0)}(p,0|1,0,1,1)\! =
	-\,\frac{1}{(4\pi)^2 \beta E}\Bigg[\left\{\frac{1}{\epsilon}-\gamma_E+\ln\left(\frac{4\pi\mu^2}{m_l^2}\right) \right\}
	\ln\sqrt{\frac{1+\beta}{1-\beta}}- \text{Li}_2\!\left(\frac{2\beta}{1+\beta}\right) 
	\nonumber\\
	&& \hspace{4.5cm} -\,  \ln^2\!\sqrt{\frac{1+\beta}{1-\beta}} 
	- i\pi\left\{\frac{1}{\epsilon}-\gamma_E+\ln\left(\frac{4\pi\mu^2}{m_l^2}\right)\right\}\Bigg]\,, \qquad\,\,
	\label{IP-1011}  
\end{eqnarray}
and
\begin{eqnarray}
	I^{+}(p^\prime,0|1,0,1,1) \!&=&\!
	\frac{1}{(4\pi)^2 \beta^\prime E^\prime}\Bigg[\left\{\frac{1}{\epsilon}-\gamma_E+\ln\left(\frac{4\pi\mu^2}{m_l^2}\right)\right\}
	\ln\sqrt{\frac{1+\beta^\prime}{1-\beta^\prime}} 
	-\text{Li}_2\!\left(\frac{2\beta^\prime}{1+\beta^\prime}\right) 
	\nonumber\\
	&& \hspace{1.6cm} -\, \ln^2\sqrt{\frac{1+\beta^\prime}{1-\beta^\prime}} 
	- i\pi\left\{\frac{1}{\epsilon}-\gamma_E+\ln\left(\frac{4\pi\mu^2}{m_l^2}\right)\right\}\,\Bigg]\,,
	\label{IP+1011_a}
\end{eqnarray}
\end{widetext}
where $\gamma_E=0.577216...$  is the Euler-Mascheroni constant.
Here, $\beta=|\vec{p}\,|/E$ and $\beta^\prime=|\vec{p^\prime}\,|/E^\prime$ are the velocities of the incoming and 
outgoing lepton, respectively, and
\begin{eqnarray}
{\rm Li}_2(z) =- \int_0^z {\rm d}t\, \frac{\ln(1-t)}{t}\,,\quad  \forall z\in {\mathbb C}\,,
\end{eqnarray}
is the standard dilogarithm (or Spence) function. For convenience we split the integral $I^{+}(p^\prime,0|1,0,1,1)$ 
into the LO [i.e., $\mathcal{O}\left(1/{M^0} \right)$] and NLO [i.e., $\mathcal{O}\left(1/M\right)$] parts, noting 
that the LO part of $I^+(p^\prime,0|1,0,1,1)$ is simply the negative of the integral $I^-(p,0|1,0,1,1)$ which contain
purely LO terms only. Thus, we have
\begin{widetext}
\begin{eqnarray}
I^+(p^\prime,0|1,0,1,1) = - I^{(0)}(p,0|1,0,1,1) + \delta^{(1/M)}I^+(p^\prime,0|1,0,1,1) + {\mathcal O}\left(M^{-2}\right)\,,
\label{IP+1011_b}    
\end{eqnarray} 
where the NLO part of $I^+(p^\prime,0|1,0,1,1)$ is given by
\begin{eqnarray}
	\delta^{(1/M)}I^+(p^\prime,0|1,0,1,1) \!&=&\! 
	-\,\frac{Q^2}{2(4\pi)^2 M E^2 \beta^3} 
	\Bigg[ \left\{\frac{1}{\epsilon}-\gamma_E+ \ln\left(\frac{4\pi\mu^2}{m_l^2}\right) \right\}\left(\ln\sqrt{\frac{1+\beta}{1-\beta}}-\beta \right) 
	-\, \text{Li}_2\!\left(\frac{2\beta}{1+\beta}\right) 
	\nonumber \\
	&&\hspace{2.6cm}+ 2 \ln\sqrt{\frac{1+\beta}{1-\beta}} - \ln^2 \sqrt{\frac{1+\beta}{1-\beta}}\,\, \Bigg]\,. 
	\label{eq:delta_I+1011}
\end{eqnarray}
\end{widetext}

Next, we have four integrals whose expressions we provide for completeness. These UV-divergent two-point integrals 
are calculated using dimensional regularization by analytically continuing the integrals to $d$-dimensional 
($d = 4 + 2\epsilon$) space-time, where the pole $\epsilon < 0$ yields the UV-divergence.\footnote{In 
Eq.~\eqref{deltaIRbox} in the main text where we isolated the IR divergence, we have analytically continued to 
dimension $D=4-2\epsilon$, where $\epsilon <0$. This case should not be confused with the continuation to the 
dimension $d\neq D$ which we use in this part of Appendix~\ref{appB} in the context of the particular UV-divergent 
integrals of interest. Nonetheless, our TPE results are free of UV-divergences, as seen in Eq.~\eqref{eq:C12inB} 
which involves taking the differences of such UV-divergent functions.} These integrals are  needed later in 
Appendix~\ref{appB} when we evaluate the tensor integrals $T^{-\mu}_1$ displayed in Appendix~\ref{appA}. They are 
given by the following expressions: 
\begin{widetext}
\begin{eqnarray}
	I^{(0)}(p,0|0,0,1,2) \!&\equiv&\! I^{-}(p,0|0,0,1,2) = \frac{1}{8\pi^2} \left[\frac{1}{\epsilon}+\gamma_E
	-\ln\left(\frac{4\pi \mu^2}{m^2_l}\right)+\frac{2}{\beta}\ln\sqrt{\frac{1+\beta}{1-\beta}}\, \right]\,,
	\\
	I^{+}(p^\prime,0|0,0,1,2) \!&=&\! \frac{1}{8\pi^2} \left[\frac{1}{\epsilon}+\gamma_E
	-\ln\left(\frac{4\pi \mu^2}{m^2_l}\right)+\frac{2}{\beta^\prime}\ln\sqrt{\frac{1+\beta^\prime}{1-\beta^\prime}}\, \right] \, = 
	\nonumber\\
	&=&\! I^{(0)}(p,0|0,0,1,2) + \frac{Q^2}{8\pi^2 ME\beta^2}
	\left[1-\left(\frac{1-\beta^2}{\beta}\right)\ln\sqrt{\frac{1+\beta}{1-\beta}}\,\right] 
	+ {\mathcal O}\left(\frac{1}{M^2}\right)\,,
\end{eqnarray}
and
\begin{eqnarray}
	I(Q|0,1,0,2) \equiv I^{-}(p,0|0,1,0,2) \equiv I^{+}(p^\prime,0|0,1,0,2) =
	\frac{1}{8\pi^2} \left[\frac{1}{\epsilon}+\gamma_E -\ln\left(\frac{4\pi \mu^2}{m_l^2}\right) 
    - \ln\left(\frac{m_l^2 M^2}{Q^4}\right)
 \right]\,,
\end{eqnarray}
\end{widetext}
where the last integral is a function of $Q^2$ only.

Having discussed the results for all divergent master integrals used in this work, we turn our attention to 
non-divergent functions. First, we enumerate the analytical expressions of the IR-finite three- and four-point
functions $I^\pm$ and $Z^\pm$ which appear in the TPE results. These integrals, which contain one heavy baryon 
propagator each, are conveniently evaluated using Zupan's method~\cite{zupan:2002}.  Again, for the sake of 
convenience and readability, it is useful to split each of these integrals into their respective LO and NLO 
parts:
\begin{widetext}
\begin{eqnarray}
I^-(p,0|0,1,1,1) \!&=&\! I^{(0)}(p,0|0,1,1,1) + \delta^{(1/M)}I^-(p,0|0,1,1,1) 
+ {\mathcal O}\left(\frac{1}{M^2}\right)\,,
\label{eq:I-}
\\
I^+(p^\prime,0|0,1,1,1) \!&=&\! -I^{(0)}(p,0|0,1,1,1) + \delta^{(1/M)}I^+(p^\prime,0|0,1,1,1) 
+ {\mathcal O}\left(\frac{1}{M^2}\right)\,,
\label{eq:I+}
\\
Z^-(\Delta,i\sqrt{-Q^2}/2,m_l,E)\!&=&\! Z^{(0)}(\Delta,i\sqrt{-Q^2}/2,m_l,E) 
+ \delta^{(1/M)}Z^{-}(\Delta,i\sqrt{-Q^2}/2,m_l,E) + {\mathcal O}\left(\frac{1}{M^2}\right)\,,
\label{eq:Z-}
\\
Z^+(\Delta^\prime,i\sqrt{-Q^2}/2,m_l,-E^\prime) \!&=&\! -Z^{(0)}(\Delta,i\sqrt{-Q^2}/2,m_l,E) 
+ \delta^{(1/M)}Z^+(\Delta^\prime,i\sqrt{-Q^2}/2,m_l,-E^\prime) +{\mathcal O}\left(\frac{1}{M^2}\right)\,,\qquad
\label{eq:Z+}
\end{eqnarray} 
where we note that the LO parts of $I^+$ and $Z^+$ differ from the LO parts of $I^-$ and $Z^-$, namely, 
\begin{eqnarray}
I^{(0)}(p,0|0,1,1,1) =-\frac{1}{ (4\pi)^2 \beta E} \Bigg[\frac{\pi^2}{6} 
- \text{Li}_2 \left(\!\frac{2 \beta}{1+\beta}\!\right)  
- \text{Li}_2 \left(\!\frac{1+\beta}{1-\beta}\!\right) - 2\ln^2 \!\sqrt{\frac{1+\beta}{1-\beta}} 
+ 2\ln \left(-\frac{Q^2}{2 M E \beta}\right) \ln \sqrt{\frac{1+\beta}{1-\beta}}\,\, \Bigg]\,,
 \nonumber\\
\label{H7}
\end{eqnarray}
and
\begin{eqnarray}
Z^{(0)}(\Delta,i\sqrt{-Q^2}/2,m_l,E) \!&=&\! 
-\,\frac{1}{(4\pi)^2\sqrt{E^2-\Delta^2}}\Bigg[\frac{1}{2}\text{Li}_2\left(\frac{\Delta^2}{\Delta^2-m^2_l}\right)
- \frac{1}{2}\text{Li}_2\left(\frac{\Delta^2-m^2_l}{\alpha^2}\right) 
- \frac{1}{2}\text{Li}_2\left(\frac{\alpha^2}{\Delta^2-m^2_l}\right) 
\nonumber\\
&&\hspace{3cm} +\, \text{Li}_2\left(1+\frac{E(1+\beta)}{\alpha}\right) 
+ \text{Li}_2\left(1+\frac{E(1-\beta)}{\alpha}\right) \Bigg]\,,
\label{Z0}
\end{eqnarray}
respectively, by overall signs only. Whereas, the NLO terms are given by the following expressions:
\begin{eqnarray}
\delta^{(1/M)}I^-(p,0|0,1,1,1) \!&=&\!  
-\,\frac{Q^2}{(4\pi)^2 M E^2 \beta^3} \Bigg[-\frac{\pi^2}{12} - \beta 
+ \frac{1}{2}\text{Li}_2 \left(\frac{2 \beta}{1+\beta}\right) 
+ \frac{1}{2}\text{Li}_2 \left(\frac{1+\beta}{1-\beta}\right) - (1+\beta) \ln \sqrt{\frac{1+\beta}{1-\beta}}   
\nonumber \\
&&\hspace{2.7cm} +\, 2 \beta \ln \sqrt{\frac{2 \beta}{1-\beta}} + \ln^2 \sqrt{\frac{1+\beta}{1-\beta}} 
- \ln \left(-\frac{Q^2}{2 M E \beta}\right) \ln \sqrt{\frac{1+\beta}{1-\beta}} 
\nonumber\\
&&\hspace{2.7cm} +\, \beta \ln \left(-\frac{Q^2}{2 M E\beta}\right) + i \pi \beta \Bigg]\,,
\label{eq:delta_I-0111}
\end{eqnarray}
\begin{eqnarray}
\delta^{(1/M)}I^+(p^\prime ,0|0,1,1,1) \!&=&\!  -\,\frac{Q^2}{(4\pi)^2 M E^2 \beta^3} 
\Bigg[\ln \sqrt{\frac{1+\beta}{1-\beta}} -\beta \Bigg]\,,
\label{eq:delta_I+0111}
\end{eqnarray}
\begin{eqnarray}
\delta^{(1/M)}Z^{-}(\Delta,i\sqrt{-Q^2}/2,m_l,E) \!&=&\! 
-\, \frac{Q^2}{4 (4 \pi)^2 M (E^2-\Delta^2)} \Bigg[(4 \pi)^2 E Z^{(0)} 
- \sqrt{\frac{\Delta^2}{\Delta^2-m^2_l}} \ln \left(\frac{\sqrt{\Delta^2} 
+ \sqrt{\Delta^2-m^2_l}}{\sqrt{\Delta^2}-\sqrt{\Delta^2-m^2_l}}\right) 
\nonumber \\
&& \hspace{3.6cm} -\, \ln \left(\frac{m_l^2}{\Delta^2-m^2_l}\right) 
- i \pi \left(1+\frac{\sqrt{\Delta^2}+E}{\sqrt{\Delta^2-m^2_l}} \right)\Bigg]\,,
\label{deltaZ-}
\\
\nonumber\\
\delta^{(1/M)}Z^+(\Delta^\prime,i\sqrt{-Q^2}/2,m_l,-E^\prime)  \!&=&\! 
\frac{Q^2}{4 (4 \pi)^2 M(E^2-\Delta^2)} \Bigg[(4 \pi)^2 E Z^{(0)} 
+ \sqrt{\frac{\Delta^2}{\Delta^2-m^2_l}} 
\ln \left(\frac{\sqrt{\Delta^2}+\sqrt{\Delta^2-m^2_l}}{\sqrt{\Delta^2}-\sqrt{\Delta^2-m^2_l}}\right)
\nonumber\\
&& \hspace{3.2cm} -\, \frac{4}{\beta} \ln \sqrt{\frac{1+\beta}{1-\beta}} - \ln \left(\frac{m_l^2}{\Delta^2-m^2_l}\right)
+ 2\left(2+\frac{E}{\sqrt{E^2-\Delta^2}}\right)
\nonumber \\
&& \hspace{3.2cm}  \times\, \ln \left(\!\frac{\sqrt{\Delta^2-m^2_l}}{\sqrt{\Delta^2-m^2_l}-\alpha}\!\right)  
- i\pi\left(1+\frac{\sqrt{\Delta^2}+E}{\sqrt{\Delta^2-m^2_l}}\right)\Bigg]\,,\quad\,\,
\label{deltaZ+}
\end{eqnarray}
\end{widetext}
where $\alpha=-E + \sqrt{E^2-\Delta^2}$, and $\Delta^2=m_{l}^{2}-\frac{1}{4}Q^{2}$. Furthermore, in the context of 
evaluating the contributions from 
(g) and 
(h) diagrams [cf. Eqs.~\eqref{delta-g} and 
\eqref{delta-h}], we additionally need to evaluate the three-point scalar master integrals $I^{-}(p,0|0,1,1,2)$ and
$I^{+}(p^\prime,0|0,1,1,2)$, respectively, as well as the tensor integrals, $I^{-\mu}_1(p,\omega|0.1,1,2)$
and $I^{+\mu}_1(p^\prime,\omega|0,1,1,2)$, respectively. These are IR-finite functions containing two powers of the
heavy baryon propagator each. First, we tackle the scalar integrals by adopting Zupan's methodology~\cite{zupan:2002}. 
For this purpose, we need to generalize our aforementioned expression for the IR-finite functions $I^-(p,0|0,1,1,1)$ 
and $I^+(p^\prime,0|0,1,1,1)$ [see Eqs.~\eqref{eq:I-}, \eqref{eq:I+}, \eqref{H7}, \eqref{eq:delta_I-0111} and 
\eqref{eq:delta_I+0111}] into $I^-(p,\omega|0,1,1,1)$ and $I^+(p^\prime,\omega|0,1,1,1)$ by introducing an 
infinitesimal parameter $\omega$, and subsequently taking the $\omega$-derivatives evaluated in the limit 
$\omega\to 0$, namely,
\begin{widetext}
\begin{eqnarray}
I^{-}(p,0|0,1,1,2) \!&=&\! -\,\lim_{\omega\to 0}\left[\frac{\partial}{\partial \omega} I^-(p,\omega|0,1,1,1)\right]
= I^{(0)}(p,0|0,1,1,2) + \mathcal{O}\left(\frac{1}{M}\right)
\nonumber\\
&=&\! -\,\frac{4 M}{(4\pi)^2 Q^2 E\beta} \ln\sqrt{\frac{1+\beta}{1-\beta}} + \mathcal{O}\left(\frac{1}{M}\right)\,,
\label{eq:I-0112}
\end{eqnarray}
\begin{eqnarray}
I^{+}(p^\prime,0|0,1,1,2) \!&=&\! -\,\lim_{\omega\to 0}\left[\frac{\partial}{\partial \omega} I^{+}(p^\prime,\omega|0,1,1,1)\right]
\nonumber\\
&=&\! -\, I^{(0)}(p ,0|0,1,1,2) - \frac{2}{(4\pi)^2 E^2 \beta^3} \left[\ln \sqrt{\frac{1+\beta}{1-\beta}}-\beta\right] 
+ \mathcal{O}\left(\frac{1}{M}\right)\,,
\label{eq:I+0112}
\end{eqnarray}
where from Eq.~(\ref{eq:generic_scalar}) we have
\begin{eqnarray}
I^{-}(p,\omega|0,1,1,1) \!&=&\! -\,\frac{1}{ (4\pi)^2 \beta E} \Bigg[\frac{\pi^2}{6} 
- \text{Li}_2 \left(\frac{2 \beta}{1+\beta}\right) - \text{Li}_2 \left(\frac{1+\beta}{1-\beta}\right) 
- 2\ln^2 \sqrt{\frac{1+\beta}{1-\beta}} 
+ 2\ln \left(\omega-\frac{Q^2}{2 M E\beta}\right) \ln \sqrt{\frac{1+\beta}{1-\beta}}
\nonumber\\
&& \hspace{1.8cm} +\, \frac{Q^2}{M E \beta^2} 
\Bigg \{- \frac{\pi^2}{12} - \beta + \frac{1}{2}\text{Li}_2 \left(\frac{2 \beta}{1+\beta}\right) 
+ \frac{1}{2}\text{Li}_2 \left(\frac{1+\beta}{1-\beta}\right) - (1+\beta) \ln \sqrt{\frac{1+\beta}{1-\beta}}
\nonumber\\
&& \hspace{3.4cm} +\, 2\beta \ln \sqrt{\frac{2 \beta}{1-\beta}} + \ln^2 \sqrt{\frac{1+\beta}{1-\beta}}  
- \ln \left(\omega-\frac{Q^2}{2 M E\beta}\right) \ln \sqrt{\frac{1+\beta}{1-\beta}} 
\nonumber\\
&& \hspace{3.4cm} +\, \beta \ln \left(\omega-\frac{Q^2}{2 M E \beta}\right) + i \pi\beta \Bigg \} 
+ \frac{2\omega}{\beta E} \Bigg\{1-\frac{1}{\beta} \ln \sqrt{\frac{1+\beta}{1-\beta}}\, \Bigg\}\, \Bigg] 
+ \mathcal{O}\left(\frac{1}{M^2}\right)\,, 
\nonumber \\
\end{eqnarray}
\begin{eqnarray}
I^{+}(p^\prime,\omega|0,1,1,1) & = &\frac{1}{ (4\pi)^2 \beta E} \Bigg[\frac{\pi^2}{6} 
- \text{Li}_2 \left(\frac{2 \beta}{1+\beta}\right) - \text{Li}_2 \left(\frac{1+\beta}{1-\beta}\right) 
- 2\ln^2 \sqrt{\frac{1+\beta}{1-\beta}} 
+ 2\ln \left(\omega-\frac{Q^2}{2 M E \beta}\right) \ln \sqrt{\frac{1+\beta}{1-\beta}}
\nonumber\\
&&\hspace{1.6cm}  +\, \frac{Q^2}{M E \beta^2} \Bigg ( \beta - \ln \sqrt{\frac{1+\beta}{1-\beta}} \Bigg ) 
- \frac{2\omega}{\beta E} \Bigg( 1 - \frac{1}{\beta} \ln \sqrt{\frac{1+\beta}{1-\beta}} \Bigg) \Bigg]
+ \mathcal{O}\left(\frac{1}{M^2}\right)\,.
\end{eqnarray}
\end{widetext}

Next, we deal with the rank-1 tensor integrals, $T^{-\mu}_1(p,0|0.1,1,2)$ and $T^{+\mu}_1(p^\prime,0|0,1,1,2)$ 
as displayed in Appendix~\ref{appA}. Using PV reduction~\cite{Passarino:1979} technique to decompose these tensor
functions into the corresponding scalar forms, we first express the integrals as follows: 
\begin{eqnarray}
T^{-\mu}_1(p,0|0,1,1,2) \!&=&\! v^\mu C^{-}_1 + p^{\prime\, \mu} C^{-}_2\,,
\nonumber\\
T^{+\mu}_1(p^\prime,0|0,1,1,2) \!&=&\! v^\mu C^{+}_1 + p^{\mu} C^{+}_2\,, 
\end{eqnarray}
noting that $T^-_1$ and $T^+_1$ becomes a function of $p^\prime$ and $p$, respectively, after performing the 
transformation of the loop momentum variable $k\to k+Q$. The above coefficients $C^\pm_{1,2}$ are then obtained 
by successively contracting with three independent available four-vectors, such as $v^\mu,\, p^\mu$ and 
$p^{\prime\mu}$, namely, $v\cdot T^\pm_1$, $p^\prime \cdot T^-_1$ and $p \cdot T^+_1$, and subsequently using IBP
to decompose these dot products into combinations of two- and three-point scalar master integrals, as discussed in
this appendix. We then obtain the following:
\begin{widetext}
\begin{eqnarray}
C^{-}_1 \!\!&=&\!\! \left[1-\frac{1}{\beta^{\prime2}}\right]\!I^{-}(p,0|0,1,1,1)  
- \left[E - \frac{E}{\beta^{\prime 2}} + \frac{m^2_l}{E^\prime \beta^{\prime 2}}\right]\!I^{-}(p,0|0,1,1,2) 
+ \frac{1}{2E^\prime\beta^{\prime 2}}\!\left[I^{-}(p,0|0,0,1,2) - I(Q|0,1,0,2)\right],
\nonumber\\
\nonumber\\
C^{-}_2 \!\!&=&\!\! \frac{1}{E^\prime\beta^{\prime 2}}I^{-}(p,0|0,1,1,1) 
- \left[\frac{E}{E^\prime\beta^{\prime 2}} - \frac{m^2_l}{E^{\prime 2} \beta^{\prime 2}}\right]I^{-}(p,0|0,1,1,2) 
- \frac{1}{2E^{\prime 2}\beta^{\prime 2}}\left[I^{-}(p,0|0,0,1,2) - I(Q|0,1,0,2)\right],
\nonumber\\
\nonumber\\
C^{+}_1 \!\!&=&\!\! \left[1-\frac{1}{\beta^2}\right]\!I^{+}(p^\prime,0|0,1,1,1) 
+ \left[E^\prime - \frac{E^\prime}{\beta^2} + \frac{m^2_l}{E \beta^2}\right]\!I^{+}(p^\prime,0|0,1,1,2) 
- \frac{1}{2E\beta^2}\!\left[I^{+}(p^\prime,0|0,0,1,2) - I(Q|0,1,0,2)\right],
\nonumber\\
\nonumber\\
C^{+}_2 \!\!&=&\!\! \frac{1}{E\beta^2}I^{+}(p^\prime,0|0,1,1,1) 
+ \left[\frac{E^\prime}{E\beta^2} - \frac{m^2_l}{E^2 \beta^2}\right]I^{+}(p^\prime,0|0,1,1,2) 
+ \frac{1}{2E^2\beta^2}\left[I^{+}(p^\prime,0|0,0,1,2) - I(Q|0,1,0,2)\right].
\label{eq:C12inB}
\end{eqnarray}  

Furthermore, there appears another finite three-point integral with one heavy baryon propagator arising from 
all but the (g) and (h) TPE diagrams, and is given by  
\begin{eqnarray}
I(Q|1,1,0,1) \!&\equiv&\! I^-(p,0|1,1,0,1) \equiv I^+(p^\prime,0|1,1,0,1) = - \frac{1}{16} \sqrt{\frac{1}{-Q^2}} 
+ {\mathcal O}\left(\frac{1}{M^2}\right)\,.
\label{I(1101)} 
\end{eqnarray}

Finally, in contrast to the aforementioned loop integrals containing the non-relativistic proton propagators, there 
exists two relativistic finite integrals essentially contributing to the seagull diagrams, namely, the scalar 
three-point function:
\begin{eqnarray}
I(Q|1,1,1,0) \equiv I^-(p,0|1,1,1,0) \equiv I^+(p^\prime,0|1,1,1,0)
=\frac{1}{8\pi^2Q^2\nu}\Bigg[\frac{\pi^2}{3}+\ln^2\sqrt{\frac{\nu+1}{\nu-1}}
+\text{Li}_2\left(\frac{\nu-1}{\nu+1}\right)\Bigg]\,, 
\label{I1110}
\end{eqnarray} 
where $\nu=\sqrt{1-4m^2_l/Q^2}$, and the tensor three-point function which can be decomposed into the following 
form: 
\begin{eqnarray}
I^{-\mu}_1(p,0|1,1,1,0) = -\frac{1}{8\pi^2Q^2\nu^2}\Bigg[\left(p^\mu-\frac{1}{2}Q^\mu\right)\ln\left(-\frac{Q^2}{m^2_l}\right)
-8\pi^2\left(Q^2p^\mu-2m^2_l Q^\mu\right)I(Q|1,1,1,0)\Bigg]\,.
  \label{seagull-tensor}
\end{eqnarray}
\end{widetext}


\bibliographystyle{apsrev}

\begin{thebibliography}{111}
\bibitem{Hofstadter:1955ae}
R.~Hofstadter and R.~W.~McAllister, 
\newblock Phys.\ Rev. \textbf{98}, 217 (1955).

\bibitem{Rosenbluth:1950yq}
M.~N.~Rosenbluth, 
\newblock Phys.\ Rev.\ \textbf{79}, 615 (1950).

\bibitem{Akhiezer:1974em}
A.~I.~Akhiezer, M.~P.~Rekalo,
\newblock Fiz.\ Elem.\ Chast.\ Atom.\ Yadra \textbf{4}, 662 (1973).

\bibitem{Arnold:1980zj}
R.~G.~Arnold and C.~E.~Carlson and F.~Gross, 
\newblock Phys.\ Rev.\ C \textbf{23}, 363 (1981).

\bibitem{Gayou:2001qt}
O.~Gayou, \textit{ et al.},  
\newblock Phys.\ Rev.\ C \textbf{64}, 038202 (2001).

\bibitem{Jones:1999rz}
M.~K.~Jones \textit{ et al,}, 
\newblock Phys.\ Rev.\ Lett. \textbf{84}, 1398 (2000).

\bibitem{Perdrisat:2006hj}
C.~F.~Perdrisat, V.~Punjabi, M.~Vanderhaeghen, 
\newblock Prog.\ Part.\ Nucl.\ Phys. \textbf{59}, 694 (2007).

\bibitem{Punjabi:2015bba}
V.~Punjabi \textit{ et al.}, 
\newblock Eur.\ Phys.\ J.\ A \textbf{51}, 79 (2015).

\bibitem{Puckett:2010}
A.~J.~R.~Puckett, \textit{ et al.}, 
\newblock Phys.\ Rev.\ Lett. \textbf{104}, 242301 (2010). 

\bibitem{Arrington:2003}
J.~Arrington,
\newblock Phys.\ Rev.\ C \textbf{68}, 034325 (2003).

\bibitem{Guichon:2003}
P.~A.~M.~Guichon and M.~Vanderhaeghen, 
\newblock Phys.\ Rev.\ Lett. \textbf{91}, 142303 (2003).

\bibitem{Blunden:2003sp}
P.~G.~Blunden, W.~Melnitchouk and J.~A.~Tjon, 
\newblock Phys.\ Rev.\ Lett. {\bf 91}, 142304 (2003).

\bibitem{Rekalo:2004wa}
M.~P.~Rekalo and E.~Tomasi-Gustafsson,
\newblock Nucl.\ Phys.\ A \textbf{742}, 322 (2004).

\bibitem{Blunden:2005ew}
P.~G.~Blunden, W.~Melnitchouk, and J.~A.~Tjon, 
\newblock Phys.\ Rev.\ C \textbf{72}, 034612 (2005).

\bibitem{Carlson:2007sp}
C.~E.~Carlson and M.~Vanderhaeghen, 
\newblock Ann.\ Rev.\ Nucl.\ Part.\ Sci. {\bf 57}, 171 (2007).

\bibitem{Arrington:2011}
J.~Arrington, P.~G.~Blunden, and W.~Melnitchouk, 
\newblock Prog.\ Part.\ Nucl.\ Phys. \textbf{66}, 782 (2011).

\bibitem{Pohl:2010zza}
R.~Pohl \textit{ et~al.}, [CREMA Collaboration] 
\newblock Nature \textbf{466} 213, 2010.

\bibitem{Pohl:2013}
R.~Pohl, R.~Gilman, G.~A.~Miller, and K.~Pachucki, 
\newblock Ann.\ Rev.\ Nucl.\ Part.\ Sci. \textbf{63}, 175 (2013).

\bibitem{Mohr:2012tt}
P.~J.~Mohr, \textit{ et al.},  
\newblock ``CODATA Recommended Values of the Fundamental Physical Constants: 2010*", 
\newblock Rev.\ Mod.\ Phys. \textbf{84}, 1527 (2012).

\bibitem{Antognini:1900ns}
A.~Antognini, \textit{ et~al.}, 
\newblock Science \textbf{339}, 417 (2013).

\bibitem{Bernauer:2014}
J.~C.~Bernauer and R.~Pohl, 
\newblock Scientific American\ \textbf{310}, 32 (2014).

\bibitem{Carlson:2015}
C.~E.~Carlson,
\newblock Prog.\ Part.\ Nucl.\ Phys. \textbf{82}, 59 (2015).

\bibitem{Bernauer:2020ont}
J.~C.~Bernauer,
\newblock EPJ\ Web\ Conf. \textbf{234}, 01001 (2020).

\bibitem{Gao:2021sml}
H.~Gao and M.~Vanderhaeghen,
\newblock Rev.\ Mod.\ Phys. \textbf{94}, 015002, (2022).

\bibitem{Kivel:2012vs}
N.~Kivel and M.~Vanderhaeghen,
\newblock JHEP \textbf{04}, 029 (2013).

\bibitem{Tomalak:2014dja}
O.~Tomalak and M.~Vanderhaeghen,
\newblock Phys.\ Rev.\ D {\bf 90}, 013006 (2014).

\bibitem{Lorenz:2014yda}
I.T.~Lorenz, U.-G.~Meissner,H.W.~Hammer and Y.B.~Dong, 
\newblock Phys.\ Rev. D {\bf 91}, 014023 (2015).

\bibitem{Tomalak:2014sva}
O.~Tomalak and M.~Vanderhaeghen, 
\newblock Eur.\ Phys.\ J.\ A {\bf 51}, 24 (2015).

\bibitem{Tomalak:2015aoa}
O.~Tomalak and M.~Vanderhaeghen,
\newblock Phys.\ Rev.\ D {\bf 93}, 013023 (2016).

\bibitem{Tomalak:2015hva}
O.~Tomalak and M.~Vanderhaeghen,
\newblock Eur.\ Phys.\ J.\ C {\bf 76}, 125 (2016).   

\bibitem{Tomalak:2016vbf}
O.~Tomalak, B.~Pasquini, M.~Vanderhaeghen, 
\newblock Phys.\ Rev.\ D \textbf{95}, 096001 (2017).

\bibitem{Tomalak:2017npu}
O.~Tomalak, 
\newblock Eur.\ Phys.\ J.\ C {\bf 77}, 858 (2017).

\bibitem{Koshchii:2017dzr}
O.~Koshchii and A.~Afanasev,
\newblock Phys.\ Rev.\ D \textbf{96}, 016005 (2017).

\bibitem{Tomalak:2018jak}
O.~Tomalak and M.~Vanderhaeghen,
\newblock Eur.\ Phys.\ J.\ C {\bf 78}, 514 (2018).

\bibitem{Talukdar:2019dko}
P.~Talukdar, V.~C.~Shastry, U.~Raha and F.~Myhrer,
\newblock Phys.\ Rev.\ D \textbf{101}, 013008 (2020).

\bibitem{Peset:2021iul}
C.~Peset, A.~Pineda and O.~Tomalak,
\newblock Prog.\ Part.\ Nucl.\ Phys. \textbf{121}, 103901 (2021).

\bibitem{Talukdar:2020aui}
P.~Talukdar, V.~C.~Shastry, U.~Raha, and F.~Myhrer,
\newblock Phys.\ Rev.\ D \textbf{104}, 053001 (2021).

\bibitem{Kaiser:2022pso}
N.~Kaiser, Y.-H.~Lin, and U.-G.~Meissner,
\newblock Phys.\ Rev.\ D \textbf{105}, 076006 (2022).

\bibitem{Guo:2022kfo}
Q.~Q.~Guo and H.~Q.~Zhou,
\newblock Phys.\ Rev.\ C \textbf{106}, 015203 (2022).

\bibitem{Gilman:2013eiv}
R.~Gilman \textit{ et~al.}, [MUSE Collaboration]
\newblock AIP\ Conf.\ Proc. \textbf{1563}, 167 (2013).

\bibitem{Gasser:1982ap}
J.~Gasser and H.~Leutwyler,
\newblock Phys.\ Rept. \textbf{87}, 77 (1982).

\bibitem{Jenkins:1990jv}
E.~Jenkins and A.~V.~Manohar, 
\newblock Phys.\ Lett.\ B \textbf{255}, 558 (1991).

\bibitem{Bernard:1992qa}
V.~Bernard, N.~Kaiser and U.-G.~Meissner,
\newblock Nucl.\ Phys.\ B \textbf{338}, 315 (1992).

\bibitem{Ecker:1994pi}
G.~Ecker,
\newblock Phys.\ Lett.\ B \textbf{336}, 508 (1994).

\bibitem{Bernard:1995dp}
V.~Bernard, N.~Kaiser and U.-G.~Meissner,
\newblock Int.\ J. Mod.\ Phys.\ E \textbf{4}, 193 (1995).

\bibitem{Cao:2021nhm}
X.~H.~Cao, Q.~Z.~Li and H.~Q.~Zheng,
\newblock Phys.\ Rev.\ D \textbf{105}, 094008 (2022).

\bibitem{Kondratyuk:2005kk}
S.~Kondratyuk, P.~G.~Blunden, W.~Melnitchouk and J.~A.~Tjon,
\newblock Phys.\ Rev.\ Lett. \textbf{95}, 172503 (2005).

\bibitem{Christy:2021snt}
M.~E.~Christy, \textit{et al.},
Phys.\ Rev.\ Lett. \textbf{128}, 102002 (2022).

\bibitem{Tsai:1961zz}
Y.~S.~Tsai,
\newblock Phys.\ Rev. \textbf{122}, 1898 (1961).

\bibitem{zupan:2002}
J.~Zupan,
\newblock Eur.\ Phys.\ J.\ C {\bf 25} (2002) 233. 

\bibitem{Chetyrkin:1981qh}
K.~G.~Chetyrkin and F.~V.~Tkachov,
\newblock Nucl.\ Phys.\ B \textbf{192}, 159 (1981).

\bibitem{Grozin:2000cm}
A.~G.~Grozin,
[arXiv:hep-ph/0008300 [hep-ph]].

\bibitem{Mo:1968cg}
L.~W.~Mo and Yung-Su Tsai.
\newblock Rev.\ Mod.\ Phys. \textbf{41}, 205 (1969).

\bibitem{Maximon:1969nw}
L.~.C.~Maximon,
\newblock Rev.\ Mod.\ Phys. \textbf{41}, 193 (1969).

\bibitem{Maximon:2000hm}
L.~C.~Maximon and J.~A.~Tjon,
\newblock Phys.\ Rev.\ C {\bf 62}, 054320 (2000).

\bibitem{McKinley:1948zz}
W.~A.~McKinley and H.~Feshbach,
\newblock Phys.\ Rev.\ \textbf{74}, 1759 (1948).

\bibitem{Kaiser:2010zz}
N.~Kaiser, 37, 115005 
\newblock J.\ Phys.\ G: Nucl.\ Part.\ Phys. \textbf{37}, 115005 (2010).

\bibitem{Passarino:1979}
G.~Passarino and M.~Veltman, 
\newblock Nucl. Phys.\ B \textbf{160}, 151 (1979).

\end{thebibliography}

\end{document}